\documentstyle[12pt,epsfig]{article}

\newlength{\dinwidth}                                                    
\newlength{\dinmargin}
\def\lapproxeq{\lower .7ex\hbox{$\;\stackrel{\textstyle                                                    
<}{\sim}\;$}}                                                    
\def\gapproxeq{\lower .7ex\hbox{$\;\stackrel{\textstyle                                                    
>}{\sim}\;$}}                                                    
\def\be{\begin{equation}}                                                    
\def\ee{\end{equation}}                                                    
\def\bea{\begin{eqnarray}}                                                    
\def\eea{\end{eqnarray}}

\begin{document}                                                    
\titlepage                                                    
\begin{flushright}                                                    
IPPP/08/92   \\
DCPT/08/184 \\                                                    
\today \\                                                    
\end{flushright}                                                    
                                                    
\vspace*{2cm}                                                    
                                                    
\begin{center}                                                    
{\Large \bf Soft processes at the LHC}\\   
\vspace*{0.5cm}
{\Large \bf I: Multi-component model}                                                                                                        
                                                    
\vspace*{1cm}                                                    
M.G. Ryskin$^{a,b}$, A.D. Martin$^a$ and V.A. Khoze$^{a,b}$ \\                                                    
                                                   
\vspace*{0.5cm}                                                    
$^a$ Institute for Particle Physics Phenomenology, University of Durham, Durham, DH1 3LE \\                                                   
$^b$ Petersburg Nuclear Physics Institute, Gatchina, St.~Petersburg, 188300, Russia            
\end{center}                                                    
                                                    
\vspace*{2cm}                                                    
                                                    
\begin{abstract}                                                    
We emphasize the sizeable effects of absorption on high-energy `soft' processes, and, hence, the necessity to include multi-Pomeron-Pomeron interactions in the usual multi-channel eikonal description.
We present a model which includes a complete set of the multi-Pomeron vertices and which accounts for the diffusion in both, the impact
parameter and $\ln(k_t)$, of the parton during its evolution
in rapidity. We tune the model to the available data for soft processes in the CERN-ISR to Tevatron energy range. We make predictions for `soft' observables at the LHC. 

\end{abstract}    
%\newpage           

\section{Motivation}

There are three main reasons for revisiting soft $pp$ high energy interactions at this time.

{\bf A.} ~This paper is concerned with the description of the high energy behaviour of ``soft'' observables such as $\sigma_{\rm tot}, ~d\sigma_{\rm el}/dt, ~d\sigma_{\rm SD}/dtdM^2,$ particle multiplicities etc. in terms of basic physics. This physics predated QCD and is sometimes regarded as the Dark Age of strong interactions. However, it is unfair to call this the Dark Age. We had a successful description of these processes in terms of the exchange of Regge trajectories linked to particle states in the crossed channels \cite{reviews}. The dominant exchange at high energy is the Pomeron, and we have Gribov's Reggeon calculus \cite{RFT} to account for the multi-Pomeron contributions. However the available data did not reach high enough energy to distinguish between the different scenarios \cite{GM1,GM2} for the high-energy
behaviour of the interaction amplitude \cite{MGR,KMR-08}.

In the `weak coupling' scenario the total cross section
$\sigma_{\rm tot}(s\to\infty)\to const$, and in order not to violate  unitarity, and
to satisfy
%provide
  the inequality
\begin{equation}
\sigma_{\rm SD}=\int\frac{d\sigma_{\rm SD}}{dM^2}dM^2\, <\, \sigma_{\rm tot},
\label{eq:1}
\end{equation}
  the triple-Pomeron vertex must vanish when $t\to 0$, that is the triple-Pomeron coupling
$g_{3P}\propto t$. In this case, the large logarithm coming from the
integration
  over the mass of the system produced in diffractive dissociation ($\int
dM^2/M^2\simeq \ln s$)
is compensated by the small value of the mean momentum transferred
through the Pomeron, $\langle t \rangle \propto 1/\ln s$.

On the other hand, in the `strong coupling' scenario, where $\sigma_{\rm tot}\propto (\ln s)^\eta$ with
$0<\eta\leq 2$, the inequality (\ref{eq:1}) is provided by a small value of the rapidity
gap survival factor $S^2$ which decreases with energy.

The present diffractive data are better described within the `strong
coupling' approach \cite{LKMR}, and in this paper we shall give predictions for the LHC for this scenario. However, the possibility of the `weak coupling' scenario is
not completely excluded yet.
Therefore, it is quite important
to study the different channels of diffractive dissociation
at the LHC in order to reach
a final conclusion and to fix the parameters of the model for
high-energy soft interactions. So the first motivation is the intrinsic interest in obtaining a reliable, self-consistent model for soft interactions, which may be illuminated by data from the LHC  \cite{early}. 

{\bf B.} ~ In turn, obtaining a reliable model will be of great value for predictions of the gross features of soft interactions. In particular, it is essential for understanding the structure of the underlying events at the LHC.

{\bf C.} ~ The third reason for studying soft interactions arises because it may not be an easy task to identify the production of
a new object at the LHC when it is accompanied by hundreds of other
particles emitted in the same event.
For
the detailed study of the new object, $A$, it may be better to select the few,
very clean, events with the Large Rapidity Gaps (LRG) on either
side of the new object, see, for example, \cite{KMRpr,KMR,DKMOR,FP420}. That is to observe the exclusive process
$pp \to p + A + p$.  In such a Central Exclusive
  Process (CEP) the mass of $A$ can be measured with very good accuracy
($\Delta M_A\sim 1-2$ GeV) by the missing-mass method by detecting the
outgoing very forward protons.  Moreover, a specific $J_z=0$ selection
rule \cite{Jz} reduces the background and also greatly simplifies the spin-parity analysis of $A$.
However, the CEP cross section is strongly suppressed by the small
survival factor, $S^2 \ll 1$, of the rapidity gaps. Thus we need a reliable model of soft interactions to evaluate
the corresponding value of $S^2$ \cite{oldsoft}. Moreover, it is important to have a model which contains $t$-channel  components of different size in order to
evaluate the possible effects of the `soft-hard factorisation' 
breaking. This is the subject of the following paper \cite{RMK2}.

\section{R\'{e}sum\'{e} of the eikonal formalism}

\subsection{Single-channel eikonal model}

First, we briefly recall the relevant features of the single-channel eikonal model. That is, we focus on elastic
unitarity. 
At high energy the position of the fast particle in the impact parameter, $b$, plane is to a good approximation frozen during the interaction, since the value of $b$ is fixed by the orbital angular momentum $l=b\sqrt{s}/2$ of the incoming hadron. There is no mixture between the partial wave amplitudes with different $l$. The well known solution of the elastic unitarity equation,
\be 2 {\rm Im}\,T_{\rm el}(s,b) = |T_{\rm el}(s,b)|^2 + G_{\rm
inel}(s,b),
 \label{eq:a1} 
\ee
may be written in terms of the phase shift $\delta_l$ as
\begin{equation}
 S_{\rm el}~ \equiv ~1+iT_{\rm el}~=~e^{2i\delta_l},~~~{\rm that~is}~~~T_{\rm el}=i(1-e^{2i\delta_l}).
\end{equation}
The presence of inelastic channels, given by $G_{\rm inel}$ in (\ref{eq:a1}), leads to the phase $\delta_l$ having an imaginary part. That is, $\delta_l$ becomes a complex number. Moreover, at high energies we know that Re$T_{\rm el}/$Im$T_{\rm el}$ is small.

\begin{figure}
\begin{center}
\includegraphics[height=1.5cm]{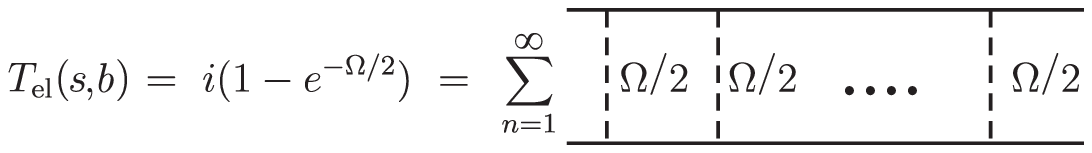}
\caption[*]{The eikonal model of elastic scattering}
\label{fig:AA}
\end{center}
\end{figure}

Now, in the framework of the eikonal model, the elastic amplitude,
\be
T_{\rm el}~=~i(1-e^{-\Omega/2})
\ee
is obtained by the sum of Regge-exchange diagrams, which is equivalent to the iteration of the elastic unitarity equation, (\ref{eq:a1}), as shown in Fig.~\ref{fig:AA}. In other words, $s$-channel elastic unitarity gives
\bea {\rm Im}T_{\rm el}(s,b) & = & 1-{\rm e}^{-\Omega/2} \label{eq:elastamp}\\
\sigma_{\rm el}(s,b) & = & (1-{\rm e}^{-\Omega/2})^2, \label{eq:el}\\
\sigma_{\rm inel}(s,b) & = & 1- {\rm e}^{-\Omega}, \label{eq:inel} \eea
where $\Omega(s,b)\geq 0$ is called the
opacity (optical density) or eikonal\footnote{Sometimes $\Omega/2$
is called the eikonal.}; $\Omega/2$ plays the role of $-2i\delta_l$. It is the Fourier transform of the two-particle ($s$-channel)
%individual Regge pole $i=R$ or $P$ takes the form
{\it irreducible} amplitude, $A(s,q_t)$. That is\footnote{We use the bold face symbols ${\mathbf q}_t$ and ${\mathbf b}$ to denote vectors in the transverse plane.}
\begin{equation}
\label{eq:append9}
\Omega (s, b) ~=~\frac{-i}{4\pi^2}\int d^2
q_t~A(s,q_t)e^{i{\mathbf q}_t\cdot{\mathbf b}} \; ,
\end{equation}
 where $q_t^2=-t$, and where the amplitude is normalized by the relation
$\sigma_{\rm tot}(s)={\rm Im}T_{\rm el}(s,t=0)$.
From (\ref{eq:inel}), we see that $ \exp(-\Omega(s,b))$ is the probability that
no inelastic scattering occurs at impact parameter $b$. 

After summing over $b$ (that is, all partial waves) we obtain the total, elastic and inelastic cross sections
\bea \sigma_{\rm tot} & = & 2\int d^2b\, {\rm Im}\,T_{\rm el}(s,b)~=~2\int d^2b~(1-{\rm e}^{-\Omega/2}) \label{eq:ot} \\
\sigma_{\rm el} & = & \int d^2b\,|T_{\rm el}(s,b)|^2 ~=~\int d^2b~(1-{\rm e}^{-\Omega/2})^2\\
\sigma_{\rm inel} & = & \int d^2b\,\left[2{\rm Im}\,T_{\rm
el}(s,b) - |T_{\rm el}(s,b)|^2\right]~=~\int d^2b~(1-{\rm e}^{-\Omega}). \eea

  Below we neglect the imaginary part of
$\Omega$, apart from the contribution of secondary Reggeons to
high-mass diffractive dissociation. At high energies, the ratio Re$T_{\rm el}/{\rm Im}T_{\rm el}$ is small,
and can be evaluated via a dispersion relation.

\subsection{Inclusion of low-mass diffractive dissociation}
So much for elastic diffraction. Now we turn to inelastic
diffraction, which is a consequence of the {\em internal
structure} of hadrons. Besides the pure elastic two-particle intermediate states shown in Fig.~\ref{fig:AA}, there is the possibility of proton excitation, $p \to N^*$. As a rule such excitations are not included in the opacity $\Omega$, but are treated separately.

This is simplest to describe at high
energies, where the lifetime of the fluctuations of the fast proton is
large, $\tau\sim E/m^2$, and during these time intervals the
corresponding Fock states can be considered as `frozen'. Each constituent of the proton can undergo scattering and thus destroy the
coherence of the fluctuations. As a consequence, the outgoing
superposition of states will be different from the incident
particle, and will most likely contain multiparticle states, so we
will have {\em inelastic}, as well as elastic, diffraction.

To discuss inelastic diffraction, it is convenient to follow Good
and Walker~\cite{GW}, and to introduce states $\phi_k$ which
diagonalize the $T$ matrix. Such eigenstates only undergo elastic
scattering. Since there are no off-diagonal transitions,
\be \langle \phi_j|T|\phi_k\rangle = 0\qquad{\rm for}\ j\neq k, 
\ee
a state $k$ cannot diffractively dissociate into a state $j$. We
have noted that this is not, in general, true for hadronic states, which are not eigenstates of the $S$-matrix, that is of $T$. To account for the internal structure of the hadronic states, we have to enlarge the set of
intermediate states, from just the single elastic channel, and to
introduce a multichannel eikonal. We will consider such an example
below, but first let us express the cross section in terms of the
probability amplitudes $F_k$ of the hadronic process proceeding via the
various diffractive eigenstates\footnote{The exponent exp$(-\Omega_k)$ describes the probability that the diffractive eigenstate $\phi_k$ is not absorbed in the interaction. Later we will see that the rapidity gap survival factors, $S^2$, can be described in terms of such eikonal exponents.} $\phi_k$.

Let us denote the orthogonal matrix which diagonalizes ${\rm
Im}\,T$ by $a$, so that
\be \label{eq:b3} {\rm Im}\,T \; = \; aT^Da^T \quad\quad {\rm with}
\quad\quad \langle \phi_j |T^D| \phi_k \rangle \; = \; T^D_k \:
\delta_{jk}. 
\label{eq:diag}
\ee
Now consider the diffractive dissociation of an arbitrary incoming
state
\be \label{eq:b4} | j \rangle \; = \; \sum_k \: a_{jk} \: | \phi_k
\rangle. \ee
The elastic scattering amplitude for this state satisfies
\be \label{eq:b5} \langle j |{\rm Im}~T| j \rangle \; = \; \sum_k
\: |a_{jk}|^2 \: T^D_k \; = \; \langle T^D \rangle, \ee
where $T^D_k \equiv \langle \phi_k |T^D| \phi_k \rangle$ and where the
brackets of $\langle T^D \rangle$ mean that we take the average of
$T^D$ over the initial probability distribution of diffractive
eigenstates. After the diffractive scattering described by
$T_{fj}$, the final state $| f \rangle$ will, in general, be a
different superposition of eigenstates from that of $| j \rangle$,
which was shown in~(\ref{eq:b4}). At high energies we may neglect
the real parts of the diffractive amplitudes. Then, for cross
sections at a given impact parameter $b$, we have
\bea \label{eq:b6} \frac{d \sigma_{\rm tot}}{d^2 b} & = & 2 \:
{\rm Im} \langle j |T| j \rangle \; = \; 2 \: \sum_k
\: |a_{jk}|^2 \: T^D_k \; = \; 2 \langle T^D \rangle \nonumber \\
& & \nonumber \\
\frac{d \sigma_{\rm el}}{d^2 b} & = & \left | \langle j |T| j
\rangle \right |^2 \; = \; \left (
\sum_k \: |a_{jk}|^2 \: T^D_k \right )^2 \; = \; \langle T^D \rangle^2 \\
& & \nonumber \\
\frac{d \sigma_{\rm el \: + \: SD}}{d^2 b} & = & \sum_k \: \left |
\langle \phi_k |T| j \rangle \right |^2 \; = \; \sum_k \:
|a_{jk}|^2 \: (T^D_k)^2 \; = \; \langle (T^D)^2 \rangle. \nonumber \eea
It follows that the cross section for the single diffractive
dissociation of a proton,
\be \label{eq:b7} \frac{d \sigma_{\rm SD}}{d^2 b} \; = \; \langle
(T^D)^2 \rangle \: - \: \langle T^D \rangle^2, \ee
is given by the statistical dispersion in the absorption
probabilities of the diffractive eigenstates. Here the average is
taken over the components $k$ of the incoming proton which
dissociates. If the averages are taken over the components of both
of the incoming particles, then in (\ref{eq:b7}) we must introduce a second index on $T^D$, that is $T^D_{ik}$, and sum over $k$ and $i$. In this case the sum is the
cross section for single and double dissociation.

At first sight, enlarging the number of eigenstates $|\phi_i\rangle$ we may include even high-mass proton dissociation. However here we face the problem of double counting when the partons originating from dissociation of the beam and `target' initial protons overlap in rapidities. For this reason high-mass dissociation is usually described by ``enhanced'' multi-Pomeron diagrams. The first and simplest is the triple-Pomeron graph, see Fig.~\ref{fig:2lad} below.

\section{Triple-Regge analysis accounting for absorptive effects}
The total and elastic proton-proton cross sections
  are usually described in terms of an eikonal model, which automatically
satisfies  $s$-channel
elastic unitarity. To account for the possibility of excitation of the initial proton, 
that is for two-particle intermediate states with the
proton replaced by $N^*$, we use the Good-Walker
formalism \cite{GW}. Already at Tevatron energies the absorptive
correction to the elastic amplitude, due to elastic eikonal
rescattering, is not negligible; it is about $-20$\% in comparison with the
simple one Pomeron exchange. After accounting for low-mass proton
excitations (that is $N^*$'s in the intermediate states) the correction becomes twice
larger (that is, about $-40$\%). Indeed, the possibility of proton excitation means that we have to include additional inelastic channels which were not accounted for in the irredicible amplitude $A$ of (\ref{eq:append9}).  This enlarges the probability of absorption for the elastic channel, that is the effective opacity $\Omega$. In terms of the Good-Walker formalism, the stronger absorption follows from the inequality\footnote{When we go from a single- to a many-channel eikonal, we may write $\Omega_k=\langle\Omega\rangle+\delta_k$ with $\langle\delta_k \rangle =0$. It follows that $\langle\Omega_k   e^{-\Omega_k}\rangle~=~\langle\Omega\rangle  \langle e^{-\Omega_k}\rangle+\langle\delta_k e^{-\delta_k}\rangle  e^{-\langle\Omega \rangle}$ which, since the second term is negative, is less than $\langle\Omega\rangle \langle e^{-\Omega_k}\rangle$.}

\be
\langle\Omega_ke^{-\Omega_k}\rangle~<~\langle\Omega_k\rangle \langle e^{-\Omega_k}\rangle,
\ee
where we average over the diffractive (Good-Walker) eigenstates.

Next, in order to describe high-mass diffractive
dissociation, $d\sigma_{\rm SD}/dM^2$, we have to include an extra factor
of 2  from the AGK cutting rules \cite{AGK}. Thus, the absorptive effects
in the triple-Regge domain are expected to be quite large. The previous
triple-Regge analyses (see, for example, \cite{FF}) did not allow for
absorptive corrections and the resulting triple-Regge couplings must be
regarded, not as bare vertices, but as effective couplings
embodying the absorptive effects \cite{capella}. Since  the inelastic cross section
(and, therefore, the absorptive corrections) expected at the LHC are more
than twice as large as that observed at fixed-target and CERN-ISR
energies,  the old triple-Regge vertices cannot be used to predict the
diffractive cross sections at the LHC.

Thus, it is necessary  to perform a new triple-Regge analysis
that includes the absorptive effects explicitly.
Such an analysis has recently been performed \cite{LKMR} using the fixed-target FNAL, CERN-ISR and Tevatron data that are
available in the triple-Regge region. The `$PPP$', `$PPR$', `$RRP$', `$RRR$' and $\pi\pi P$ contributions,
were included assuming either the `strong' or `weak' coupling scenarios for the
behaviour of the triple-Pomeron vertex. To account for the absorptive
corrections a two-channel (Good-Walker) eikonal model was used, which describes well the
total, $\sigma_{\rm tot}$, and elastic, $d\sigma_{\rm el}/dt$, $pp$ and $\bar
pp$ cross sections.

In the `strong' coupling  case, a good
$\chi^2/$DoF=167/(210-8)=0.83 was obtained.  In
comparison with the old triple-Regge analysis \cite{FF}, a
twice larger relative contribution of the `$PPR$' term was found. This is mainly due
to the inclusion of the higher-energy
Tevatron data in the analysis.

Since the absorptive effects are included explicitly,
   the extracted values of the triple-Reggeon vertices are now much closer to the {\it
bare} triple-Regge couplings. In particular, the value
\begin{equation}
g_{PPP}\, \equiv \, \lambda g_N,~~~~~~~~~{\rm where}~~\lambda \simeq 0.2
\label{eq:2}
\end{equation}
is consistent with a reasonable extrapolation of the perturbative BFKL
Pomeron vertex to the low scale region \cite{BRV}; here $g_N$ is the Pomeron-proton coupling.  Note also that these
values of the `$PPP$' and `$PPR$' vertices allow a good description of the HERA
data \cite{Jpsi} on inelastic $J/\psi$ photoproduction, $\gamma p\to
J/\psi+Y$, where the screening corrections are rather small.

The `weak' coupling scenario leads to a larger $\chi^2/$DoF=1.4 and to a
worse description of the $\gamma p\to J/\psi+Y$ process at the lowest
values of $t$. At the LHC energy the `weak' coupling fit predicts about 3 times smaller inclusive cross section
$d\sigma_{\rm SD}/dtdM^2$ at $\xi=M^2/s=0.01$ and low $t$ in comparison with that predicted in the `strong'
coupling case.

\section{Model with a complete set of multi-Pomeron vertices}

Note that the effects due to the triple-Pomeron vertex (\ref{eq:2}) are rather large. Indeed, the contribution caused by such vertices is enhanced by the logarithmically large phase space available in rapidity. In particular, the total cross section of high-mass dissociation is roughly\footnote{Here, for simplicity, we assume an essentially flat energy dependence, $\sigma \sim s^{\epsilon}$ with $\epsilon {\rm ln}s <1$.} of the form
\be
\sigma_{\rm SD}~=~\int \frac{M^2 d\sigma_{\rm SD}}{dM^2}~\frac{dM^2}{M^2}~\sim~\lambda{\rm ln}s~\sigma_{\rm el},
\ee
where $\lambda$ reflects the suppression of high-mass dissociation in comparison with elastic scattering and the ln$s$ factor comes from the integration $\int dM^2/M^2~\sim~{\rm ln}s$. Thus actually we deal with the parameter $\lambda {\rm ln}s \gapproxeq 1$ at collider energies. For each fixed rapidity interval the probability of high-mass dissociation (or, in other words, the contribution due to the triple-Pomeron vertex) is relatively small. However the cumulative effect in the complete interaction amplitude is {\rm enhanced} by the large phase space available in rapidity.

\begin{figure}
\begin{center}
\includegraphics[height=3cm]{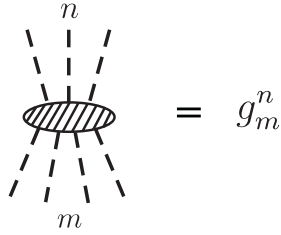}
\caption[*]{A multi-Pomeron vertex}
\label{fig:gnm}
\end{center}
\end{figure}

As a consequence, the contribution of the corresponding, so-called `enhanced', diagrams, with a few
vertices,
  is not negligible. Moreover, we cannot expect that more complicated
multi-Pomeron interactions, driven by the $g^n_m$ vertices, which
describe the transition of $n$ to $m$ Pomerons of Fig.~\ref{fig:gnm}, will not affect the
final result.  It  looks more reasonable to assume that $g^n_m\propto
\lambda^{n+m}$ than to assume that $g^n_m=0$ for any $n+m>3$. Thus we
need a model which accounts for the possibility of multi-Pomeron interactions
(with arbitrary $n$ and $m$).

In this paper we extend and develop the {\it partonic} approach of Ref.~\cite{KMRs1}.
While the eikonal formalism describes the rescattering of the incoming
fast particles,  the enhanced multi-Pomeron diagrams
represent the rescattering of the intermediate partons in the ladder
(Feynman diagram) which describes the Pomeron-exchange amplitude.

\begin{figure}
\begin{center}
\includegraphics[height=3cm]{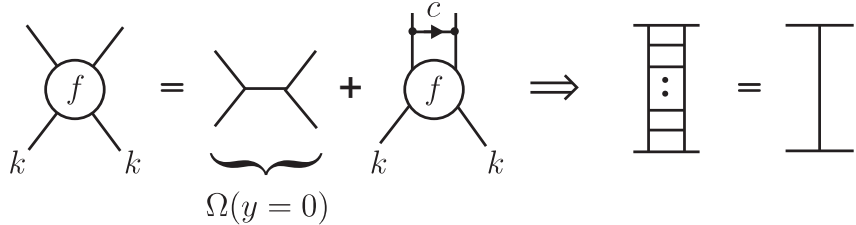}
\caption[*]{The evolution of the elastic bare Pomeron amplitude, $\Omega_k(y,b)$.}
\label{fig:dT}
\end{center}
\end{figure}
Indeed, we start with the generation of ladder-type structure of `elastic' bare Pomeron exchange amplitude. It may be generated by the evolution equation \cite{AFS}
(in rapidity, $y$, space)
\begin{equation}
\frac{d\Omega(y,b)}{dy}\,=\,
\left(\Delta+\alpha'\frac{d^2}{d^2b}\right)\Omega(y,b),
\label{eq:3}
\end{equation}
where $b$ is the two-dimensional vector in impact parameter
space. $\Delta$ is
the probability to emit new intermediate partons (denoted $c$) within unit
rapidity interval; it is analogous to the splitting function of DGLAP evolution. The impact parameter of parton $c$ is not frozen in the evolution. At each step $b$ can be changed by a constant amount $\Delta b$ in any direction, leading to the diffusion represented by the second term on the right-hand side of (\ref{eq:3}) where $\alpha'$ plays the role of the diffusion coefficient \cite{CDR}. The evolution is shown symbolically in Fig.~\ref{fig:dT}.
The solution of (\ref{eq:3}) is
\be
\Omega(y,b)=\Omega_0\exp(y\Delta -b^2/4\alpha' y)/4\pi\alpha' y.
\label{eq:yb}
\ee
It represents the opacity (at point $y,b$), corresponding to the incoming
particle placed at $b=0$ and $y=0$.

\begin{figure}
\begin{center}
\includegraphics[height=3cm]{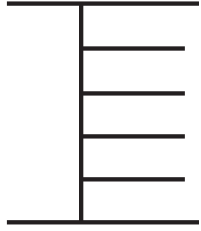}
\caption[*]{A pure inelastic high multiplicity process}
\label{fig:mult}
\end{center}
\end{figure}

It may be helpful to explain why (\ref{eq:3}) was written in terms of the opacity $\Omega$. First we note that the discontinuity of the amplitude generated by (\ref{eq:3}) does not contain a two-particle $s$-channel intermediate state; it corresponds to a pure inelastic high multiplicity process, see Fig.~\ref{fig:mult}. Due to elastic unitarity, (\ref{eq:a1}) this inelastic interaction leads to elastic $pp$ scattering. Thus we have to put the solution of (\ref{eq:3}) into the eikonal formulae of (\ref{eq:elastamp})-(\ref{eq:inel}), as it does not contain two-particle $s$-channel states.

In momentum space the solution (\ref{eq:yb}) corresponds to the amplitude
\be
A(s,t)~=~A_0~s^{1+\Delta+\alpha' t}.
\ee
That is to the  bare Pomeron exchange amplitude, where the Pomeron trajectory has intercept $\alpha(0)=1+\Delta$ and slope $\alpha'$.

\begin{figure}
\begin{center}
\includegraphics[height=3cm]{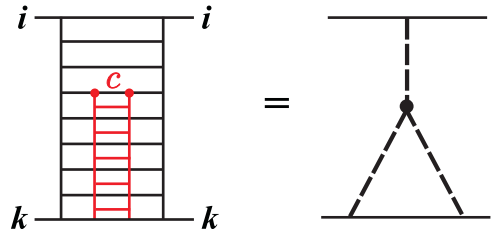}
\caption[*]{The ladder structure of the triple-Pomeron amplitude.}
\label{fig:2lad}
\end{center}
\end{figure}

A multi-Pomeron enhanced contribution arises from the absorption of intermediate $s$-channel partons $c$ during the evolution of $\Omega$ in $y$. The simplest example is the triple-Pomeron diagram in which parton $c$ undergoes an extra rescattering with the target parton $k$, as shown in Fig.~\ref{fig:2lad}. 
Allowing for many rescatterings, we have to sum over different numbers of ladders between partons $c$ and $k$. Assuming an eikonal form for the multi-Pomeron-proton vertex, it is natural to replace (\ref{eq:3}) by
\be
\frac{d\Omega_k}{dy}~=~e^{-\lambda\Omega_k/2}\left(\Delta+\alpha'\frac{d^2}{d^2b}\right)\Omega_k(y,b)
\label{eq:Teik}
\ee
where the `opacity' $\Omega_k$ describes the transparency of the target $k$. As we are dealing with the elastic {\it amplitude} we use $e^{-\lambda\Omega_k/2}$ and not $e^{-\lambda\Omega_k}$. The coefficient $\lambda$ arises since  parton $c$ will have a different absorption cross section from that of eigenstate $i$. Naively, we may assume that the beam $i$ contains a number $1/\lambda$ of partons. The factor $e^{-\lambda\Omega_k/2}$  generates multi-Pomeron vertices of the form 
\begin{equation}
g^n_m\,=\, n\cdot m\cdot\lambda^{n+m-2}g_N/2\;\;\;\;\;\;\;\;\;
\mbox{for $n+m\geq 3$}\, .
\label{eq:g}
\end{equation}
where $g^n_m$ is defined in Fig.~\ref{fig:gnm}. Even though $\lambda \simeq 0.25$, the role of the factor $e^{-\lambda\Omega_k/2}$ is not negligible, since the suppression effect is accumulated throughout the evolution. For instance, if $\lambda \ll 1$ the full absorptive correction is given by the product $\lambda \Omega Y/2$, where the small value of $\lambda$ is compensated by the large rapidity interval $Y$. 

In terms of Regge diagrams, (\ref{eq:Teik}) sums up the system of fan diagrams in which any number $m$ of ``lower'' Pomerons couples to a fan vertex $g_m^1$.  Such multi-Pomeron diagrams are called ``enhanced'', since their contribution is enhanced, in comparison with the eikonal diagrams, by the large available rapidity interval $Y$. In order to include the rescattering with the beam $i$ we replace (\ref{eq:Teik}) by
\begin{equation}
\frac{d\Omega_k(y,b)}{dy}\,=\,e^{-\lambda(\Omega_k(y,b)+\Omega_i(y',b))/2}
\left(\Delta+\alpha'\frac{d^2}{d^2b}\right)\Omega_k(y,b)\; ,
\label{eq:4}
\end{equation}
The final term in the exponent is the opacity of the beam $i$, which depends on the rapidity interval $y'=Y-y$, with $Y={\rm ln}s$. The equation for the opacity $\Omega_i$ has an analogous form
\begin{equation}
\frac{d\Omega_i(y',b)}{dy'}\,=\, e^{-\lambda(\Omega_i(y',b)+\Omega_k(y,b))/2}
\left(\Delta+\alpha'\frac{d^2}{d^2b}\right)\Omega_i(y',b),
\label{eq:5}
\end{equation}
in which we now evolve in the opposite direction starting from the boundary condition $\Omega(y'=0)$ at $y=Y$.

Recall that the fit to the data in the triple-Reggeon domain indicated a very small
(consistent with zero) $t$-slope of all the triple-Reggeon
vertices \cite{LKMR,FF}.  Thus, as the size of the
multi-Reggeon vertices are negligible in comparison with the size of the
incoming hadron, we may write the absorptive corrections (that is, the
exponential factors on the right-hand-side  of
 (\ref{eq:4},\ref{eq:5}))
such that the opacities $\Omega_i,\, \Omega_k$ are taken at the same
point in the impact parameter plane $b$.

Since the intermediate parton may be absorbed by the
interaction
with the particles (partons) from the wave function of both the beam or target
hadron, we now need to solve the two equations, (\ref{eq:4})
and (\ref{eq:5}). This is done iteratively. Moreover, note that the opacities $\Omega$ now depend on two
 vectors in impact parameter space - the separation ${\mathbf b}_1$ between
 the position of the intermediate parton $c$ and the beam hadron, and the separation ${\mathbf b}_2$  between $c$ and the target hadron. The argument $b$
%${\mathbf b}$ 
in (\ref{eq:4},\ref{eq:5}) now symbolically denotes both
${\mathbf b}_1$ and ${\mathbf b}_2$.
The resulting solution $\Omega(y,{\mathbf b})$ is then used in the eikonal formulae
for the elastic amplitude, giving
\be
 T_{\rm el}({\mathbf b})=1-\exp(-\Omega({\mathbf b}={\mathbf b}_1-{\mathbf b}_2)/2).
\ee
A more detailed description of the amplitude, and the cross sections of the
different diffractive processes can be found in Ref.~\cite{KMRs1}.

\subsection{Multi-components in both the $s$- and $t$-channels}

As mentioned above, as in \cite{KMRs1}, we use three diffractive components in the $s$-channel. 
In other words, we use a 3-channel eikonal for the rescattering
of fast particles. The transverse size squared of each eigenstate is proportional to the corresponding absorptive
cross section; $R^2_i\propto \sigma_i$.  That is, we assume that the parton density
at the origin is the same for each eigenstate. The shape of the
Pomeron-nucleon vertex is  parametrised by the form factor
$V(t)=e^{d_2 t}/(1-t/d_1)^2$, whose Fourier transform, $V({\mathbf b})$, plays the
role of the initial conditions for $\Omega(y=0,{\mathbf b})$.  However, now, we allow for a
non-zero slope ($\alpha'\neq 0$) of the (bare) Pomeron trajectory.

A major development, of the model of \cite{KMRs1}, is that we use four different $t$-channel states, which we label $a$: one for the
secondary Reggeon ($R$) trajectory and three Pomeron states ($P_1, P_2, P_3$) to mimic the BFKL
diffusion in the logarithm of parton transverse momentum,
$\ln(k_t)$ \cite{Lip}. To be precise, since the
BFKL Pomeron \cite{bfkl} is not a pole in the complex $j$-plane, but a
branch cut,
we approximate the cut by three  $t$-channel states of a different size.
The typical values of $k_t$ in each of the three states is about $k_{t1}\sim 0.5$ GeV,
$k_{t2}\sim 1.5$ GeV and $k_{t3}\sim 5$ GeV.
Thus the system of evolution equations (\ref{eq:4},\ref{eq:5}) is replaced by\footnote{Strictly speaking both opacities
 $\Omega_i$ and $\Omega_k$ depend on both subscripts $i$ and $k$. Here we keep
 only one subscript to distinguish the parent hadron for
each active parton (gluon).}

\begin{equation}
\frac{d\Omega^a_k(y,{\mathbf b}_1,{\mathbf b}_2)}{dy}=
e^{-\lambda[\overline\Omega^a_k(y,{\mathbf b}_1,{\mathbf b}_2)+
\overline\Omega^a_i(y',{\mathbf b}_1,{\mathbf b}_2)]/2}
\left(\Delta^a+\alpha'_a\frac{d^2}{d^2b_1}\right)\Omega^a_k(y,{\mathbf b}_1,{\mathbf b}_2)
+V_{aa'}\Omega^{a'}_k,
\label{eq:6}
\end{equation}
\begin{equation}
\frac{d\Omega^a_i(y',{\mathbf b}_1,{\mathbf b}_2)}{dy'}=
e^{-\lambda[\overline
\Omega^a_k(y,{\mathbf b}_1,{\mathbf b}_2)+\overline\Omega^a_i(y',{\mathbf b}_1,{\mathbf b}_2)]/2}
\left(\Delta^a+\alpha'_a\frac{d^2}{d^2b_2}\right)\Omega^a_i(y',{\mathbf b}_1,{\mathbf b}_2)
+V_{aa'}\Omega^{a'}_i,
\label{eq:7}
\end{equation}
where $\Delta^a=\alpha(0)-1$ and $\alpha'_a=\alpha'_P$ for $a=P_1,P_2,P_3$, while for the secondary Reggeon, $(a=R)$, which is built of quarks, we take $\Delta^R=\alpha_R(0)=0.6$ and $\alpha'_R=0.9~ {\rm GeV}^{-2}$, so that the last term $V_{RR}\Omega^R$ is diagonal with $V_{RR}=-1$ to account for the spin $\frac{1}{2}$ nature of quarks.  The key parameters which drive the evolution are the intercepts $\Delta$ and the slopes $\alpha'$. In general, each component $a$ may have different values of $\Delta_a$ and $\alpha'_a$. We discuss the values in Section 5.

In the exponents, the opacities $\bar\Omega_i$
($\bar\Omega_k$) are actually the sum of the opacities
$\Omega^{a'}_i$ ($\Omega^{a'}_k$) with corresponding coefficients.
Namely
$$
\overline\Omega^{P_1}=\Omega^{P_1}+\Omega^{P_2}v_{_{PP}}+\Omega^Rv_{_{PR}}
$$
$$
\overline\Omega^{P_2}=\Omega^{P_2}+\Omega^{P_1}v_{_{PP}}+\Omega^{P_3}v'_{_{PP}}
$$
$$
\overline\Omega^{P_3}=\Omega^{P_3}+\Omega^{P_2}v'_{_{PP}}
$$
\begin{equation}
\overline\Omega^{R}=\Omega^{P_1}v_{_{RP}}+\Omega^Rv_{_{RR}}.
\label{eq:ombar}
\end{equation}
We chose $v_{_{PP}}=(1/3)^2$ since, at the leading order, the probability
of interaction of two components of different size ($k_t$)
is proportional to the ratio $(k_{t2}/k_{t1})^2$. We take
$v'_{_{PP}}=1/27$ since the third (smallest size) component collects
{\it all} the higher $k_t$ contributions, and therefore here the mean
value of $k_t$ is larger. For the screening of the secondary Reggeon by
the Pomeron we take just the colour factor
$v_{_{RP}}=C_F/C_A=(4/9)$, as we assume that the secondary Reggeon is
composed of a $t$-channel quark-antiquark pair. Finally the factors %in the last equation
$v_{PR}=1.8$ and $v_{RR}=4$
 were tuned to give a reasonable reproduction of the secondary Reggeon
contributions to the available $pp\to p+X$ data\footnote{Note that all the opacities in the absorptive exponents are multiplied by $\lambda=0.25$. Thus,  the value of the product $\lambda v_{RR}=1$ is not large.}.
%$v_{PR}=1.8$ and $v_{RR}=4$.

The transition factors $V_{aa'}$ between the different $t$-channel components are fixed by the properties of the BFKL
equation. The only non-zero factors, apart from $V_{RR}=-1$, are
\begin{equation}
 V_{P_1P_2}=\rho^{P_2}v_{_{PP}},~~~V_{P_2P_1}=\rho^{P_1}v_{_{PP}},~~~V_{P_2P_3}=\rho^{P_3}v'_{_{PP}},~~~V_{P_3P_2}=\rho^{P_2}v'_{_{PP}},
\end{equation}
where
\begin{equation}
\rho^a_{ik}=\Delta^a
e^{-\lambda({\overline\Omega^a_k(y,b)+
\overline\Omega^a_i(y',b))/2}}.
\label{eq:rho}
\end{equation}
 $\rho^a$ is the density of partons emitted in the rapidity evolution
of the $t$-channel component $a$. The remaining transition factors were set to zero. That is
\be
V_{P_3P_1}=V_{P1P_3}=0,~~~V_{Ra'}=V_{aR}=0, ~~~{\rm and}~~~ V_{aa'}=0 ~~{\rm for}~~ a=a'.
\ee

For each Good-Walker $s$-channel component ($i,k$), the initial conditions are fixed
by the parton (matter) distribution in the corresponding diffractive eigenstate
\begin{equation}
\Omega^a_i(y=0,{\mathbf b})=\frac{\beta_i({\mathbf b})\beta_0}{4\pi}=\frac{\beta_0}{4\pi^2}
\int e^{i{\mathbf q}_t \cdot {\mathbf b}}\beta_i(t)d^2q_t\; ,
\label{eq:inom}
\end{equation}
where $\beta_i(t)=\gamma_i\beta(\gamma_it)$. The parametrisation
 $\beta(t)=\beta_0e^{d_2t}/(1-t/d_1)^2$ was used for the Pomeron,
while for the secondary Reggeon we chose the Gaussian form
$\beta(t)=\beta_Re^{d_Rt}$.

The relative couplings (and the corresponding size) of the components
were taken to (a) reproduce the cross section of low-mass dissociation
measured at the CERN-ISR \cite{CERN-ISR}, and (b) to make all three components quite
different from each other; $\gamma_1=1.80$,
$\gamma_2=0.82$,
$\gamma_3=0.38$.
All the coefficients in the decomposition $|p\rangle =\sum a_i|\phi_i\rangle$
are taken to be $a_i=1/\sqrt 3$ (with $i=1,2,3$).

To avoid possible double counting, and to fix the boundary
between the `low' and `high' mass dissociation, we introduce a threshold
$\Delta y=1.5$ in rapidity for the actual start of the evolution of
(\ref{eq:6},\ref{eq:7}). That is we start the evolution at $y=\Delta y=1.5$ and not at $y=0$. Hence
the available rapidity interval becomes $\delta Y=\ln(s)-2\Delta y$.
Proton excitation (dissociation) which covers a rapidity interval
larger than $\Delta y$ (i.e. $\ln(M^2/s_0)>1.5$) is called `high-mass dissociation'.

It is natural to separate the different contributions in terms of rapidity, since in QCD the interference between the different diagrams for gluon radiation leads to angular (rapidity) ordering of emitted gluons, at least to leading log accuracy\footnote{Therefore it will be interesting and important to measure the single diffractive cross section, not only in the usual form $d\sigma_{\rm SD}/dM^2$, but also in the form $d\sigma_ {\rm SD}/d\eta$, where $\eta$ denotes the position of the edge of the rapidity gap. This may be possible using the forward shower counters proposed in \cite{FSC}.}. 

Note that the initial condition (\ref{eq:inom}) is only valid for the
secondary Reggeon and for the large size Pomeron component ($a=P_1$). For the smaller size Pomeron components we use
\be
\Omega^{P_2}_i(y=0,{\mathbf b})=\Omega^{P_1}_i(y=0,{\mathbf b})v_{_{PP}}~~{\rm and}~~
\Omega^{P_3}_i(y=0,{\mathbf b})=\Omega^{P_1}_i(y=0,{\mathbf b})
v_{_{PP}}v_{_{PP}}'.
\ee

\subsection{Total and differential cross section formulae}
\begin{figure}
\begin{center}
\includegraphics[height=4cm]{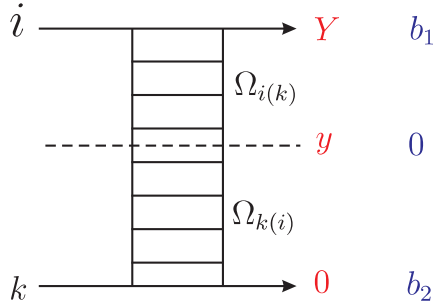}
\caption[*]{The irreducible amplitude $F_{ik}(Y,{\mathbf b})$ of a high energy interaction.}
\label{fig:Fik}
\end{center}
\end{figure}
To calculate the elastic amplitude we need the $s$-channel two-particle {\it irreducible} amplitudes for the scattering of the various diffractive eigenstates $i$ and $k$, for given separations  ${\mathbf b}={\mathbf b}_1-{\mathbf b}_2$
 between the incoming protons. These are given by
  \begin{equation}
 F_{ik}(Y,{\mathbf b})=\frac
 1{\beta_0^2}\sum_a\int\Omega^a_{ik}(y,{\mathbf b}_1,{\mathbf b}_2)\Omega^a_{ik}(Y-y,{\mathbf b}_1,{\mathbf b}_2)
 d^2b_1d^2b_2\delta^{(2)}({\mathbf b}_1-{\mathbf b}_2-{\mathbf b})
 \label{eq:F}
 \end{equation}
 where $Y=\ln s$, see Fig.~\ref{fig:Fik}. Note that there is no integral\footnote{The integral over $y$ gives the multiplicity.} over $y$. The convolution may be calculated at any rapidity $y$, leading to the same result. Given this effective `$ik$ eikonal', we can calculate the cross sections (analogously to (\ref{eq:elastamp})-(\ref{eq:inel})). We obtain
  \begin{equation}
 \sigma_{\rm tot}(Y,{\mathbf b})=2\sum_{i,k}|a_i|^2|a_k|^2\int\left(1-e^{F_{ik}(Y,{\mathbf b})/2}
 \right) d^2b \; ,
 \label{eq:st}
 \end{equation}
\begin{equation}
 \frac{d\sigma_{\rm el}(Y,{\mathbf b})}{dt}=\frac
 1{4\pi}\left[\int d^2be^{i  {\mathbf q}_t\cdot {\mathbf b}}\sum_{i,k}|a_i|^2|a_k|^2
 \left(1-e^{F_{ik}(Y,{\mathbf b})/2} \right)\right]^2
 \label{eq:dsel}
 \end{equation}
 where $t=-q^2_t$, and
  \begin{equation}
 \sigma_{\rm el}(Y,{\mathbf b})=\int d^2b\left[\sum_{i,k}|a_i|^2|a_k|^2
 \int\left(1-e^{F_{ik}(Y,{\mathbf b})/2} \right)\right]^2  \; .
 \label{eq:sel}
 \end{equation}

\subsection{Low-mass diffractive dissociation}

\begin{figure}
\begin{center}
\includegraphics[height=2cm]{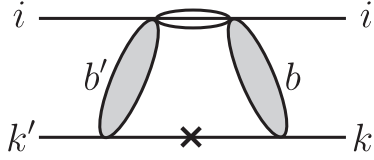}
\caption[*]{The symbolic diagram for (\ref{eq:delsd}) for low-mass dissociation of the `beam' diffractive eigenstate $i$.}
\label{fig:elSD}
\end{center}
\end{figure}

For low-mass excitation of the beam proton we obtain
 $$\frac{d\sigma_{\rm el+SD}(Y,{\mathbf b})}{dt}=\frac
 1{4\pi}\sum_{i,k,k'}|a_i|^2|a_k|^2|a_{k'}|^2~~~ \times $$
  \begin{equation}
 ~~~~~~\times~~~\left[\int d^2b'e^{-i{\mathbf q}_t\cdot{\mathbf b}'}
 \left(1-e^{F_{ik'}(Y,{\mathbf b}')/2} \right)\right]
 \left[\int d^2be^{i{\mathbf q}_t\cdot{\mathbf b}}
 \left(1-e^{F_{ik}(Y,{\mathbf b})/2} \right)\right]\; ,
 \label{eq:delsd}
 \end{equation}
which has the symbolic structure shown in Fig.~\ref{fig:elSD}.

Strictly speaking, we may need a different diagonalisation matrix $a$ of (\ref{eq:diag}) for the different $t$-channel exchanges. However, if the main difference between the diffractive eigenstates is due to the size  and the impact parameter structure of the state, which is frozen for a fast hadron during the interaction, then it is justified to use the same eigenstates for any $t$-channel exchange, $\Omega^a$.

\subsection{High-mass diffractive dissociation}

The expression for the high-mass excitation is more complicated.
The cross section for beam particle diffractive dissociation (with the
  gap up to $y$) can be written using (\ref{eq:elastamp})-(\ref{eq:inel}).
 Diffractive dissociation may be considered as the
 elastic scattering of intermediate parton $c$ caused by its
absorption on the target, which is described by the factor
$\exp(-\lambda\overline\Omega_k/2)$.

%$$ \sigma_t=2\int d^2b(1-e^{\Omega({\mathbf b})/2}),\;\sigma_{el}=\int d^2b(1-
%e^{-\Omega({\mathbf b})/2})^2\;\; \mbox{and}\;\; \sigma_{inel}=\int
%d^2b(1-e^{-\Omega({\mathbf b})})$$

\begin{figure}
\begin{center}
\includegraphics[height=3cm]{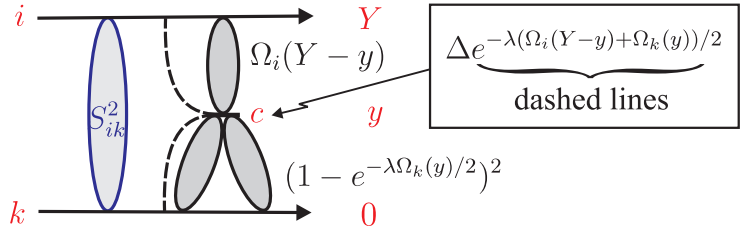}
\caption[*]{The symbolic diagram for (\ref{eq:sd}) for high-mass dissociation of the `beam' diffractive eigenstate $i$.}
\label{fig:SDM}
\end{center}
\end{figure}

Thus, in each impact parameter point ${\mathbf b}$ the cross section for  single
dissociation is proportional to (i) the elastic $c-k$ cross section
$(1-\exp(-\lambda\overline
\Omega_k(y,{\mathbf b})/2))^2$; (ii) to the probability to find
the intermediate parton $c$ in the interval $dy$, that is
$\Delta\exp(-\lambda\overline
\Omega_i/2-\lambda\overline\Omega_k/2)$; (iii) to the
amplitude $\Omega_i$ of the parton $c$-beam interaction; (iv) to the
gap survival factor $S^2({\mathbf b})=\exp(-\Omega(Y,{\mathbf b}))$ ($Y=\ln s$). The
resulting cross section reads
$$\frac{d\sigma_{\rm SD}}{dy}\,=\, N\int
(1-e^{-\lambda\Omega_k(y,{\mathbf b}_1,{\mathbf b}_2)/2})^2\Delta
e^{-\lambda\Omega_i(Y-y,{\mathbf b}_1,{\mathbf b}_2)/2-\lambda\Omega_k(y,{\mathbf b}_1,{\mathbf b}_2)/2}~~~\times $$
\begin{equation}
\times~~~\Omega_i(Y-y,{\mathbf b}_1,{\mathbf b}_2)S^2_{ik}(|{\mathbf b}_1-{\mathbf b}_2|)
d^2b_1 d^2b_2\; ,
\label{eq:sd}
\end{equation}
where ${\mathbf b}_1$ (${\mathbf b}_2$) are the coordinates in
the impact parameter plane with respect to the beam (target) hadron.
The normalisation factor $N$ is specified in (\ref{eq:sdf}).
The gap survival probability\footnote{Strictly speaking, when calculating the
gap survival probability in each particular case, we only have to account for
the possibility of rescattering which
produces secondaries within the gap interval. That is, in
(\ref{eq:s}) we should not put the whole irreducible amplitude
$F_{ik}({\mathbf b})$, but, instead, part of it; since the contribution from the processes
with a gap in the same (or a larger) rapidity interval does not change
 qualitatively the structure of the diffractive dissociation
event. In the present computations we neglect this
effect. This means that actually the gap survival probabilities, and
the true cross sections of diffractive dissociation, should be a bit
larger.}
\begin{equation}
S^2_{ik}({\mathbf b})=\exp(-F_{ik}({\mathbf b}))\; .
\label{eq:s}
\end{equation}
The symbolic structure of (\ref{eq:sd}), for high-mass single dissociation, is shown in Fig.~\ref{fig:SDM}.

Accounting for the different Good-Walker eigenstates and the different
states in $t$-channel we obtain
%\footnote{To make the expression shorter
%we omit here the arguments ${\mathbf b}_1$ and ${\mathbf b}_2$ of the opacities $\Omega$.}
$$\frac{M^2d\sigma_{\rm SD}}{dM^2}\,=\sum_i|a_i|^2\int
\left|\sum_k|a_k|^2\sum_a T^a_{ik}(y,{\mathbf b}_1,{\mathbf b}_2)
\sqrt{\rho^a_{ik}(y,{\mathbf b}_1,{\mathbf b}_2))}S_{ik}(|{\mathbf b}_1-{\mathbf b}_2|)\right|^2~~\times
$$
\begin{equation}
\times~~\Omega^a_i(Y-y,{\mathbf b}_1,{\mathbf b}_2)d^2b_1d^2b_2/\beta_0^2\, ,
\label{eq:sdf}
\end{equation}
where the parton density $\rho^a_{ik}$ is defined by eq.(\ref{eq:rho})
and, neglecting the secondary reggeon contribution, the elastic
 $c-k$ amplitude
\begin{equation}
T^a_{ik}(y,{\mathbf b}_1,{\mathbf b}_2)=
\left(1-e^{-\lambda\overline\Omega^a_k(y,{\mathbf b}_1,{\mathbf b}_2)/2}\right)\; .
\label{eq:T}
\end{equation}

For the secondary Reggeon, the real part may be large: note that $\alpha_{f_2}(\langle t\rangle) \sim 0.2$. To allow for this, and since, that besides the $f_2$-trajectory with vacuum quantum numbers, there exists also $\omega,~\rho$ and $a_2$ exchange, we enlarge the contribution due to $\Omega^R$ by increasing the values of the effective triple-Reggeon couplings $g_{PPR},~g_{RRP},~g_{RRR}$ as compared to those coming from the absorptive opacities (\ref{eq:ombar}).
 So we use (\ref{eq:T}) for $a=P_2$ or $P_3$. On the other hand, for $a=P_1$ or $R$ we use, respectively,
\begin{equation} T^{P_1}_{ik}(y,{\mathbf b}_1,{\mathbf b}_2)=
\sqrt{\left(1-e^{-\lambda\overline\Omega^{P_1}_k(y,{\mathbf b}_1,{\mathbf b}_2)/2}\right)^2
+r_{RRP}\left(1-e^{-\lambda\Omega^{R}_k(y,{\mathbf b}_1,{\mathbf b}_2)v_{_{PR}}/2}\right)^2}\; .
\label{eq:T1}
\end{equation}
and
\begin{equation}
T^{R}_{ik}(y,{\mathbf b}_1,{\mathbf b}_2)=
\sqrt{\left(1-e^{-\lambda\Omega'_k(y,{\mathbf b}_1,{\mathbf b}_2)/2}\right)^2
+r_{RRR}\left(1-e^{-\lambda\Omega^{R}_k(y,{\mathbf b}_1,{\mathbf b}_2) v_{_{RR}}/2}\right)^2}\; .
\label{eq:TR}
\end{equation}
where $\Omega'=r_{PPR}\Omega^{P_1}+\Omega^Rv_{RR}$. We take $r_{RRP}=r_{PPR}=3,\; r_{RRR}=9$ to reproduce the available data in the CERN-ISR to Tevatron energy range.

The slope of the diffractive dissociation cross section,
$B_{\rm SD}=d[\ln(d\sigma_{\rm SD}/dM^2]/dt$ at $t=0$, can be calculated as
the mean value of ${\mathbf b}^2_2$ -- the separation of the intermediate parton
$c$ from the target hadron
$$B_{\rm SD}=\sum_i|a_i|^2\int
\left|\sum_k|a_k|^2\sum_a T^a_{ik}(y,{\mathbf b}_1,{\mathbf b}_2)
\sqrt{\rho^a_{ik}(y,{\mathbf b}_1,{\mathbf b}_2))}S_{ik}(|{\mathbf b}_1-{\mathbf b}_2|)\right|^2~~\times
$$
\begin{equation}
~~\times~~ {\mathbf b}^2_2\Omega^a_i(Y-y,{\mathbf b}_1,{\mathbf b}_2)\frac{d^2b_1d^2b_2}{\beta_0^2}
\left[\frac{M^2d\sigma_{\rm SD}}{dM^2}\right]^{-1}\, .
% \label{eq:sdf}
\end{equation}

\subsection{Central exclusive production}

Central Exclusive Diffractive (CED) production of a  system with mass squared $M^2=\xi_1\xi_2 s$
with the large rapidity gaps either side, which is sometimes called the Double-Pomeron-Exchange (DPE) process, has a cross section given by
\begin{equation}
\frac{\xi_1\xi_2d\sigma_{\rm CED}}{d\xi_1d\xi_2}\ =\ \sum_{a,a'}
\int\left|\sum_{i,k} E^{a'}_i E^a_k \Omega^{aa'}(y_1,y_2,{\mathbf b}'_1,{\mathbf b}_2) S_{ik}\right|^2 d^2b_1 d^2b_2 d^2b'_2/\beta^2_0,
\label{eq:ced}
\end{equation}
where
\begin{equation}
E^{a'}_i\ =\ |a_i|^2 T^{a'}_{ik}(Y-y_1,{\mathbf b}_1,{\mathbf b}'_2)\sqrt{\rho^{a'}_{ik}(Y-y_1,{\mathbf b}_1,{\mathbf b}'_2)}\, ,
\end{equation}
\begin{equation}
E^{a}_k\ =\ |a_k|^2 T^{a}_{ki}(y_2,{\mathbf b}'_1,{\mathbf b}_2)\sqrt{\rho^{a}_{ik}(y_2,{\mathbf b}'_1,{\mathbf b}_2)}
\end{equation}
are the probability amplitudes for elastic scattering of the 
intermediate parton $c$ ($c'$) on the beam (target) eigenstate
 $i$ ($k$). The coordinates of parton $c$ ($c'$) are ${\mathbf b}_1$ and 
${\mathbf b}'_2$  (${\mathbf b}'_1$ and ${\mathbf b}_2$) with respect to the beam and the target proton respectively; that is, ${\mathbf b}'_1={\mathbf b}_1-{\mathbf b}'_2+{\mathbf b}_2$.
The momentum fractions ($\xi_i=1-x_{L,i}$) of the incoming protons, transferred across the gaps, are $\xi_1=e^{-(Y-y_1)}$ and $\xi_2=e^{-y_2}$.
 The gap survival factor $S_{ik}(b=|{\mathbf b}_1-{\mathbf b}'_2|)$ is given
by (\ref{eq:s}).

The amplitude of the interaction of partons $c$ and $c'$, 
$\Omega^{aa'}(y_1,y_2,{\mathbf b}'_2,{\mathbf b}_2))$, is obtained by the solution of the evolution (\ref{eq:6}), which {\it starts} from the initial condition $\Omega^a(y=y_2)=\delta^{(2)}({\mathbf b}'_2-{\mathbf b}_2)$. That is, it {\it starts} from  one parton at rapidity $y_2$ placed at coordinate ${\mathbf b}_2$ in $t$-channel state $a$, and {\it finishes} at the point $y_1,{\mathbf b}'_2$ in the state $a'$; note $y_1 > y_2$. After the usual solution of (\ref{eq:6},\ref{eq:7}), the evolution (\ref{eq:6}) was performed in the known ``background'' fields $\overline{\Omega}^a_k,\overline{\Omega}^a_i$ to account for the
absorption of intermediate partons.

 \section{Description of the data and predictions for the LHC}

Clearly, the number of parameters in our model is too large to perform a straightforward $\chi^2$ fit of the data. Instead, we fix the
majority of the parameters at reasonable values and
demonstrate that such a model can reproduce all the features of the available data on
diffractive cross sections, $\sigma_{\rm tot},\, d\sigma_{\rm el}/dt,\,
\sigma_{\rm SD}^{{\rm low}M},\,
d\sigma_{\rm SD}/dM^2$. 

\begin{figure} [t] 
\begin{center}
\includegraphics[height=12cm]{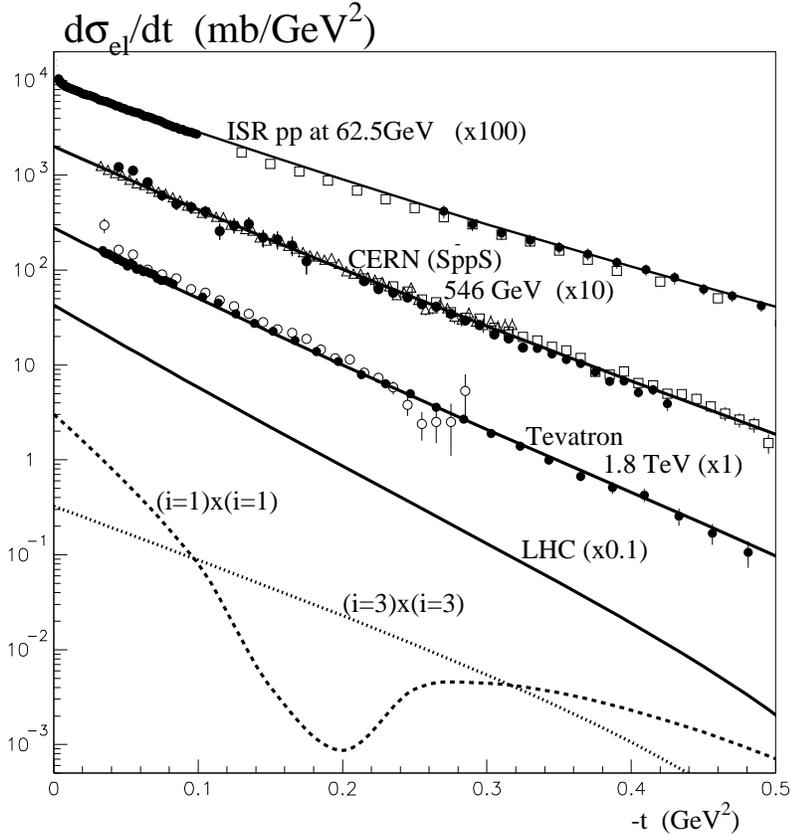}
%\epsf{figure=sumlog.eps,width=14.0cm,height=14.0cm}
%\centerline{\epsfxsize=2.9in\epsfbox{sumlog.eps}}
\caption[*]{The $t$ dependence of the elastic $pp$ cross section. The
dashed and dotted lines are the contributions from  the elastic
scattering of the largest size ($i=1$) and the smallest size ($i=3$)
components.} 
\label{fig:dsdt}
\end{center}
\end{figure}

Before we give the values of the parameters, we show in Fig.~\ref{fig:dsdt} the quality of the description of the data for the {\it elastic differential} cross section. 
We also present in Fig.~\ref{fig:dsdt} the prediction for differential elastic cross section at
the LHC energy $\sqrt{s}=14$ TeV.
Recall that we are using a three-channel eikonal. That is $i,k=1,2,3$. It is interesting to note that the contribution to the cross section arising from the scattering of the two large-size
eigenstates, $(i=1)\times(i=1)$, already has a diffractive dip at $-t=0.2$ GeV$^2$. However, after the contributions from all
possible combinations $i\times k$ are summed up, the
prediction has no dip up to $-t=0.5$ GeV$^2$. 

Note, also, that the Pomeron and secondary Reggeon couplings to the proton were taken to have the forms
\be
\beta(t)~=~\beta_0e^{d_2 t}/(1-t/d_1)^2, ~~~~~~~~~~~\beta(t)=\beta_Re^{d_R t}.
\ee 
The values of the parameters that we  use are 
 \be
d_2=0.15~ {\rm GeV}^{-2},~~~~d_1=1.5~{\rm GeV}^2,~~~~d_R=1~{\rm GeV}^{-2}.
\ee
The non-zero value of $d_2$ is simply to provide good
  convergence and accuracy of the Fourier transform. The parameter
$d_1$ controls the $t$ behaviour of the elastic cross section, while
$d_R$ is responsible for the $t$ slope of diffractive dissociation at relatively
low $y=-\ln\xi$, where the cross section is dominated by the $RRP$
triple-Reggeon term. The relative size of this contribution, as compared to that due to $PPP$, was tuned by choosing $v_{PR}=1.8$ and $r_{RRP}=3$. In order to describe the data, the couplings were found to be $\beta_0^2=33$ mb and $\beta_R^2=8$ mb. Since we choose a relatively simple $t$ dependence for the Reggeon-proton couplings $\beta(t)$, our model is only reliable over a restricted $t$ interval, $-t \lapproxeq 0.5 ~{\rm GeV}^2$. Note that in this domain, the real part of Pomeron exchange, and a possible Odderon exchange contribution, would give only very small effects.

\begin{figure} [t]
\begin{center}
\includegraphics[height=15cm]{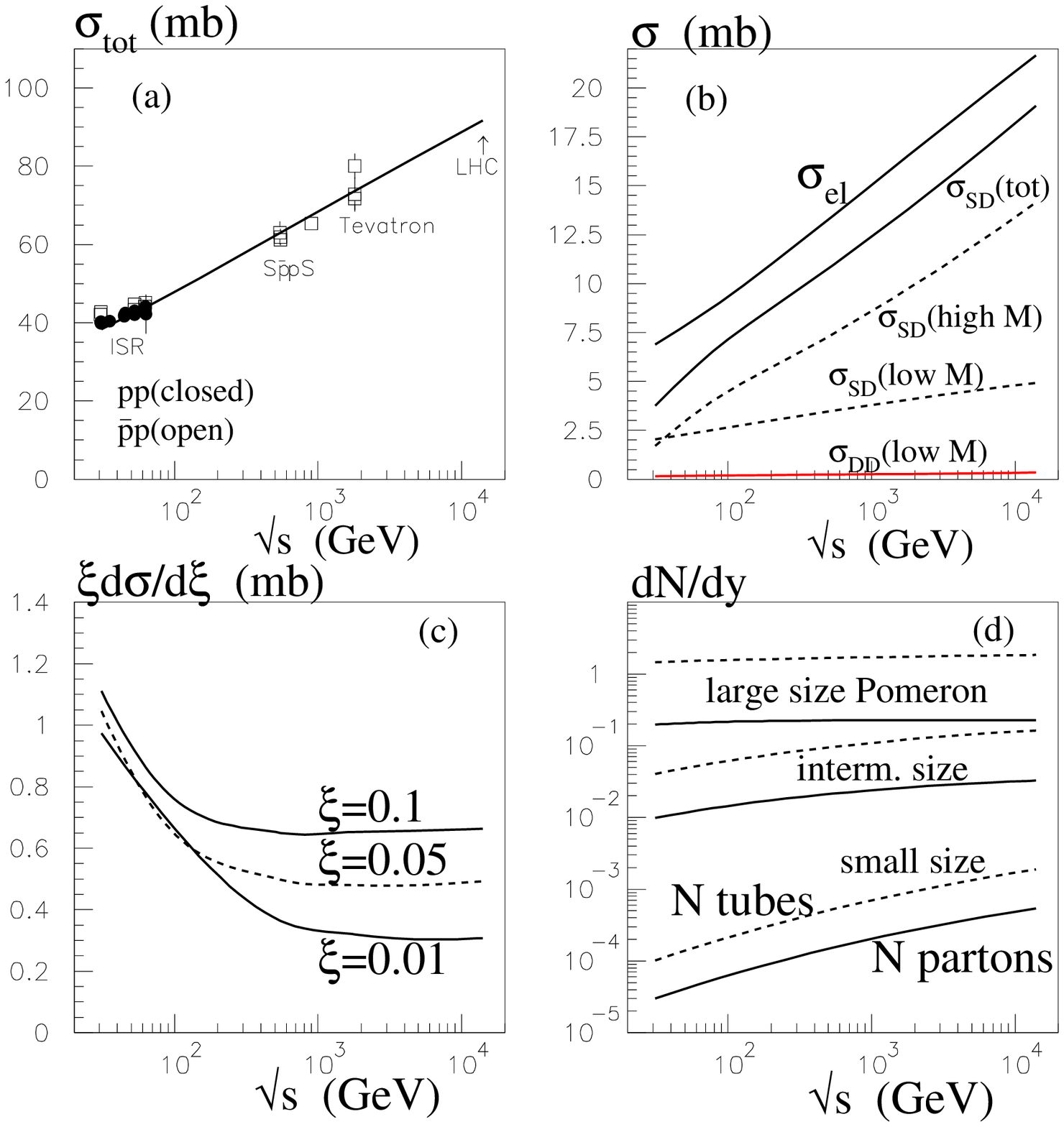}
%\epsf{figure=sumlog.eps,width=14.0cm,height=14.0cm}
%\centerline{\epsfxsize=2.9in\epsfbox{sumlog.eps}}
\caption[*]{The energy dependence of the total (a), elastic and
diffractive dissociation (b) $pp$ cross sections and the cross sections of
dissociation to a fixed $M^2=\xi s$ state (c); (d) the parton
multiplicity (solid lines) and the number of 'colour  tubes' (dashed)
produced by the Pomeron components of different size.}
\label{fig:sum4}
\end{center}
\end{figure}

To describe the high energy behavior of the {\it total} cross section, we take $\Delta^a=0.3$ for each of the three components of the Pomeron.
These Pomeron intercepts are consistent with resummed NLL BFKL,
which gives $\omega_0\sim 0.3$ practically independent of the scale
$k_t$ \cite{BFKLnnl}. The slopes of the Pomeron trajectories are driven by the transverse momentum associated with the particular component $a$. In fact, we have $\alpha'\propto 1/k^2_t$.
We find the data require $\alpha'_{P_1}=0.05$ GeV$^{-2}$ for the large-size Pomeron component, so we put
$\alpha'_{P_2}=0.05/9$ GeV$^{-2}$ for the second component and $\alpha'_{P_3}=0$ for the smallest-size component.
For the secondary Reggeon trajectory we take $\alpha'_R=0.9$ GeV$^{-2}$,
and $\alpha_R(0)=0.6$. The `bare' value is a bit larger than $\frac{1}{2}$,
since the final effective intercept is reduceed by the absorptive
corrections included in the evolution equation. The description of the total cross section data are shown in Fig.~\ref{fig:sum4}(a). The screening corrections arising from the `enhanced' multi-Pomeron
  diagrams, that is from the high-mass dissociation, slow down the growth of
  the cross section with energy.  Thus, the model predicts a  relatively low
total
cross section at the LHC -- $\,\sigma_{\rm tot}({\rm LHC})\simeq 90$
mb\footnote{This value is also predicted by other models of `soft' interactions which include absorptive effects \cite{KGB,GLMM}.}. 

\subsection{Low-mass dissociation}

Recall that the couplings of the Good-Walker eigenstates $i$ were specified by $\beta_i(t)=\gamma_i\beta(\gamma_i t)$. The values $\gamma_1=1.80$, $\gamma_2=0.82$ and
$\gamma_3=0.38$ were chosen so as to reproduce the {\it low-mass dissociation}
cross section $\sigma_{\rm SD}^{{\rm low} M}=2$ mb  at the CERN-ISR energy\footnote{Although, here, we use a three-channel eikonal model, 
practically the same results, and the same quality of the description,
is obtained using a two-channel eikonal, that is only two
eigenstates $|\phi_i\rangle$ (see also the discussion in \cite{KMRs1}).} \cite{CERN-ISR}.

\subsection{High-mass dissociation}
\begin{figure} [b]
\begin{center}
\includegraphics[height=8cm]{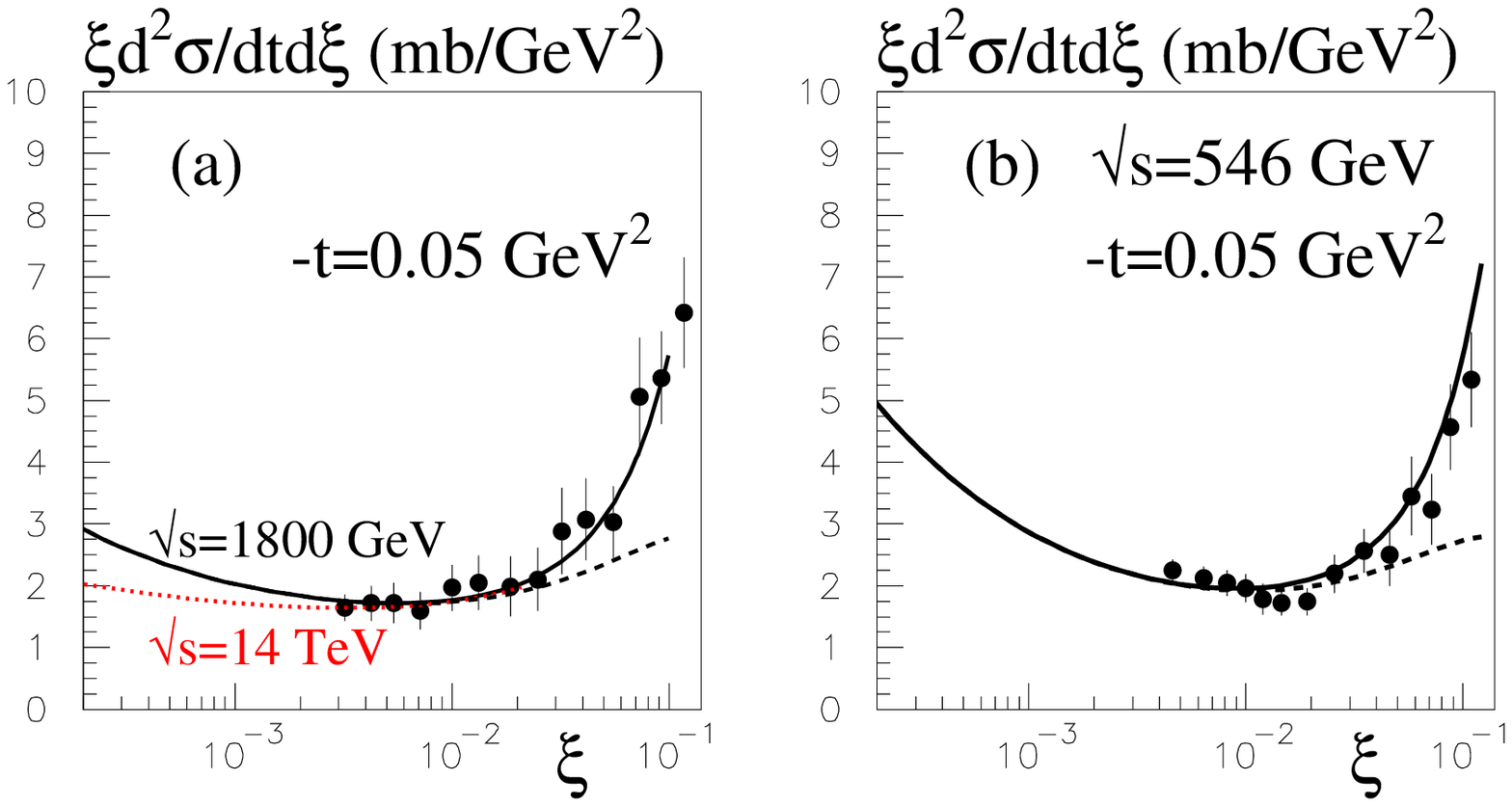}
%\epsf{figure=sumlog.eps,width=14.0cm,height=14.0cm}
%\centerline{\epsfxsize=2.9in\epsfbox{sumlog.eps}}
\caption[*]{The model description of the data for the cross section for high-mass dissociation versus $\xi$ for $-t=0.05~{\rm GeV}^2$ at $\sqrt{s}=$ 1800 GeV and 546 GeV \cite{CDFhm,GoulMon}. The dashed lines are the predictions without the $\pi\pi P$ contribution. The dotted curve at small $\xi$ is the prediction for the LHC.}
\label{fig:dsy}
\end{center}
\end{figure}

The value of the parameter $\lambda$, which controls the cross section of {\it high-mass
dissociation} in the small $\xi$ (that is, large $y$) region, was found to be $\lambda=0.25$.
The dependence of the cross section for high-mass dissociation, $\xi d^2\sigma/dtd\xi$, on $\xi=M^2/s$ is compared with the Tevatron CDF data \cite{CDFhm,GoulMon} in Fig.~\ref{fig:dsy}. Recall that in our model we have not included pion exchange.  Since the $\pi\pi P$ term is essential at  large $\xi$, we have included the corresponding contribution using the parameters obtained in \cite{LKMR}. The results without the $\pi\pi P$ term are shown by the dashed lines. We also show in Fig.~\ref{fig:dsy}(a) (by the dotted line at small $\xi$) the prediction for the LHC energy. 
% In addition to these CDF data, we also obtain a good description %of all the `triple-Regge' data used in Ref.~\cite{LKMR}.

Above, we introduced the different types of data mentioning the parameters that they mainly constrain. Of course, in practice, these parameter values are used to describe all the `soft' data simultaneously.

The energy behaviour of the cross sections are shown in Table 1 and Fig.~\ref{fig:sum4}. Fig.~\ref{fig:sum4} also shows the energy behaviour of the
multiplicities of the secondaries produced by the $t$-channel Pomeron
components of different sizes; we will discuss
the multiplicity distributions in some detail in Section 6.
\begin{table}[htb]
\begin{center}
\begin{tabular}{|l|c|c|c|c|c|}\hline
energy &   $\sigma_{\rm tot}$ &  $\sigma_{\rm el}$ &    $\sigma_{\rm SD}^{{\rm low}M}$ &  $\sigma_{\rm SD}^{{\rm high}M}$  &   $\sigma_{\rm SD}^{\rm tot}$ \\ \hline

 1.8  &   73.7   &     16.4  &       4.1      &   9.7  &     13.8  \\
 14    &  91.7   &     21.5     &    4.9   &     14.1  &     19.0    \\
 100   &  108.0  &     26.2  &       5.6   &     18.6  &     24.2   \\ \hline

\end{tabular}
\end{center}
\caption{Cross sections (in mb) versus collider energy (in TeV).}
\end{table}

The values of $\sigma_{\rm SD}^{\rm tot}$ quoted in Table 1 look, at first sight, too large, when compared with the value  $9.46 \pm 0.44$ mb given by CDF \cite{CDFhm}. However the CDF value does not include the secondary Reggeon ($RRP$) contribution, denoted as a `non-diffractive' component of $2.6 \pm 0.4$ mb. Moreover, the trigger used to select the diffractive dissociation events rejects part of the low-mass proton excitations. Taking these absences into account, there is no contradiction between the model prediction and the CDF data. Furthermore, note that in the region where the CDF detector efficiency and resolution are good, our model gives an excellent description of the measured data, see  Fig.~\ref{fig:dsy}.

It is interesting to note, that after tuning the parameters to describe all the available `soft' data, the model satisfies the Finite Energy Sum Rules \cite{DHS} to good accuracy\footnote{We thank Alan White for discussions.}. Indeed, we can switch off the low-mass dissociation, putting the same couplings for each diffractive eigenstate ($\gamma_1=\gamma_2=\gamma_3=1$), and replace the low-mass excitations for $\Delta y <1.5$ by the triple- and multi-Regge contributions\footnote{Recall that in the basic model we introduced a threshold $\Delta y=1.5$; we started the evolution (\ref{eq:6}), (\ref{eq:7}) at $y=1.5$ in order not to generate low-mass dissociation via triple- and multi-Regee contributions and to avoid double counting.}. Keeping all the other parameters as before, we then obtain $\sigma_{\rm tot}=$73 mb (93 mb) and $\sigma_{\rm SD}^{\rm tot}$=13.6 mb (20.1 mb) for the Tevatron (LHC) energies, These values are close to those in Table 1.

In principle, it is straightforward, although computer intensive, to use the model to calculate the cross section for double dissociation, $\sigma_{\rm DD}$. We do not show the values here. The values will be similar to those in Table 2 of Ref.~\cite{KMRs1}; footnote 25 of that paper shows that the values of $\sigma_{\rm DD}$ are in excellent agreement with the Tevatron data.

\subsection{Central exclusive production}
\begin{figure} [h]
\begin{center}
\includegraphics[height=10cm]{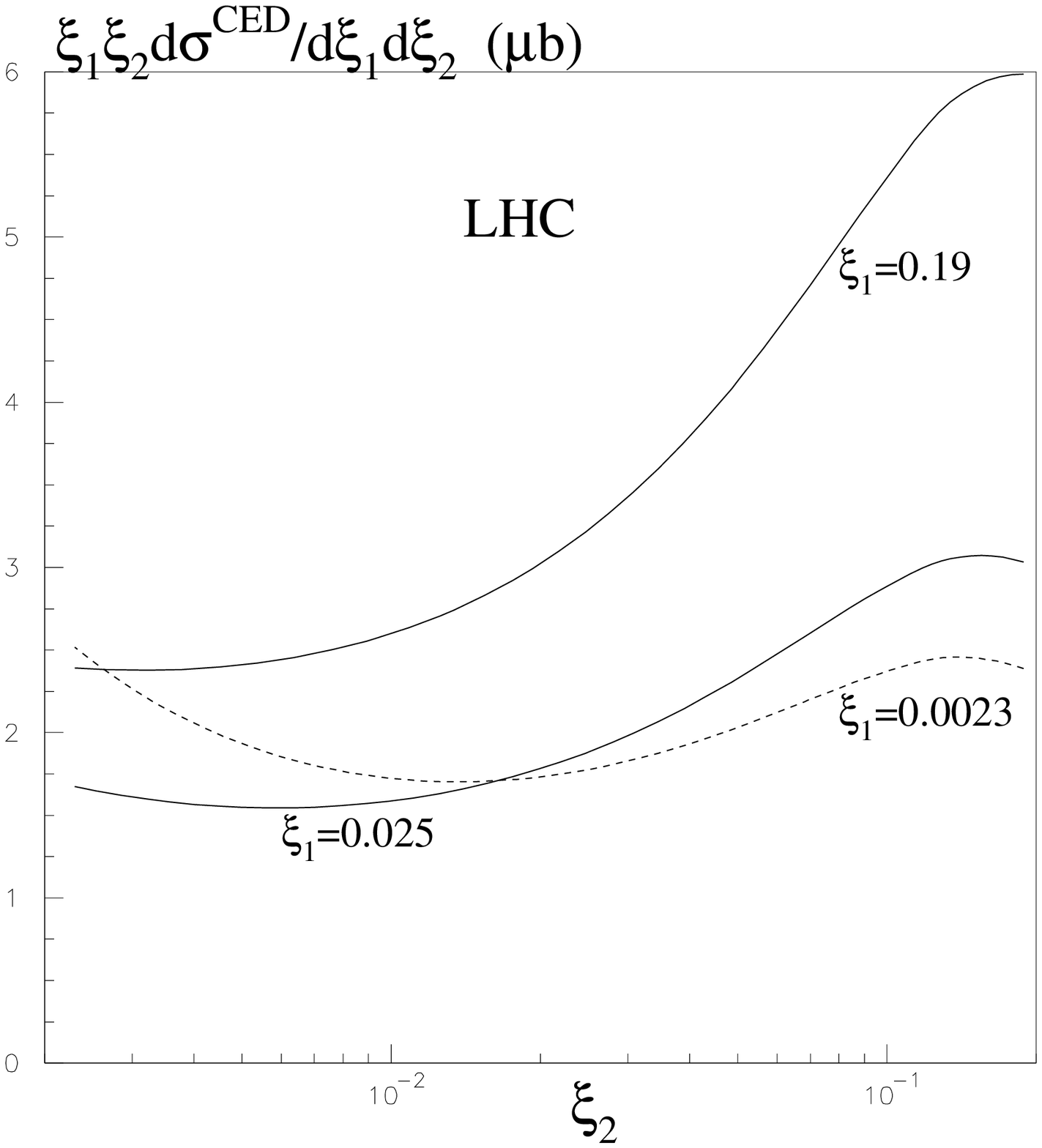}
%\epsf{figure=sumlog.eps,width=14.0cm,height=14.0cm}
%\centerline{\epsfxsize=2.9in\epsfbox{sumlog.eps}}
\caption[*]{Sample predictions for Central Exclusive Diffractive production at the LHC. The $\xi_i$'s are the momentum fractions of the incoming protons transferred across the rapidity gaps on either side of the centrally produced system of mass $M=\sqrt{\xi_1 \xi_2 s}$.}
\label{fig:ced}
\end{center}
\end{figure}

The cross sections for the Central Exclusive Diffractive (CED) production at the LHC energy $\sqrt s=14$ TeV are shown in Fig.~\ref{fig:ced} for those $\xi$ intervals which can be studied by the TOTEM and FP420 detectors. Here we mean the soft CED production of a state
with the mass given by $M^2=\xi_1\xi_2 s$ separated from the incoming protons by
two large rapidity gaps.
The calculation is described in Section 4.5.  The  cross section integrated over the $0.002<\xi_i<0.2$  ($0.02<\xi_i<0.2$) intervals (for both $\xi_1$ and $\xi_2$) is predicted\footnote{To speed up the computation we neglect
the small non-zero value of $\alpha'$ in
the calculation of the amplitude $\Omega^{aa'}$.
%for the small non-zero value of $\alpha'$.
Then there is no diffusion in impact parameter space and the integral over ${\mathbf b}'_2$ disapears; since
${\mathbf b}'_2={\mathbf b}_2$. However, to correct the final result we smear out the resulting amplitude, which allows for the larger gap survival
probability at larger $b$. In this way we retain reasonable ($\sim 20$\%) accuracy of the computations.} to be 53 (16) $\mu$b.
 The major contribution comes from  pure soft interactions. For $\xi<0.02$, more than 80\% of the cross section is due to the large size component of the Pomeron; and more than half for larger values of $\xi$.
Note that, in the CED calculations, we did not include the $\pi\pi P$ contribution. Thus, actually, the expected cross section will be larger for $\xi$ values that are not too small, see Fig.~\ref{fig:dsy}. 

Note that the resulting CED cross section is about twice larger than that expected  from the naive factorization formula
\begin{equation}
\frac{\xi_1\xi_2d\sigma_{\rm CED}}{d\xi_1d\xi_2}=\frac 1{\sigma_{\rm tot}} \frac{\xi_1d\sigma_{\rm SD}}{d\xi_1}\frac{\xi_2d\sigma_{\rm SD}}{d\xi_2}\ .
\label{eq:fac}
\end{equation}
This is due to the fact that each single dissociation contains a gap survival factor $S^2_{ik}(b)$ and therefore the r.h.s. of
(\ref{eq:fac}) is proportional to $(S^2)^2$ while the l.h.s. contains $S^2$ once only. On the other hand, for Central Exclusive Production the typical values of the impact parameter $b$ are smaller; so we have a smaller gap survival factor
$\langle S^2(b) \rangle$. This partly compensates the additional power of $S^2$ in the r.h.s. of (\ref{eq:fac}), and as a result the violation of the factorization shown in (\ref{eq:fac}) is not so strong. We will discuss rapidity gap survival in central exclusive production in detail in the following paper \cite{RMK2}.

 \section{Multiparticle inclusive production}

As we have a detailed model for high energy soft processes, it would appear to be possible to predict the multiplicity distribution at the LHC. However, although some general features can be predicted, it is not so simple to make a quantitative prediction. We address the problem below.

Recall that in the evolution equations for the {\it amplitude}, given in
(\ref{eq:6}), (\ref{eq:7}), we
include the absorptive factor $\exp(-\Omega/2)$, and not
$\exp(-\Omega)$. That is we work with the forward amplitude
  ${\rm Im}T(b)=1-e^{-\Omega/2}$, which at each step of the
  evolution (in rapidity $y$)
includes all possible processes - both elastic and inelastic
  interactions with cross sections
$\sigma_{\rm el}(b)=(1-e^{-\Omega/2})^2$ and
$\sigma_{\rm inel}(b)=1-e^{-\Omega}$; where $\sigma_{\rm tot}(b)=2{\rm Im}
T(b)=\sigma_{\rm el}(b)+\sigma_{\rm inel}(b)$, see eqs.
(\ref{eq:elastamp})-(\ref{eq:inel}).

As usual, inelastic processes include both single-ladder
exchange, as well as multiple interactions with a larger density of
secondary partons. The situation is similar to the rescattering of a fast hadron in
a heavy nucleus.  That is, in such an eikonal approach
the probability, $w_N (b)$, of events with parton multiplicity $N$ times
larger than that in a single ladder, is given by
 \begin{equation}
 w_N=\frac{\Omega^N}{N!}e^{-\Omega}.
\end{equation}
In the multi-channel case the opacity $\Omega$ should be replaced by
$F_{ik}$. Unfortunately, we cannot
use this probability $w_N$ literally to describe the
multiplicity distributions of secondary {\it hadrons}.
In particular
 a non-negligible fraction of the final hadrons may be produced via the
fragmentation of minijets. These processes are beyond the `pure soft'
approach used in the present paper. Therefore, below, we discuss
 the multiplicity distribution only at the partonic level.

 The mean number of the ($t$-channel) ladders of the type $a$
 produced in the collision of $i$ and $k$ Good-Walker eigenstates
 can be calculated as
\begin{equation}
 N^a_{ik}({\mathbf b})=\frac{1}
{\beta_0^2\sigma_{ik}({\mathbf b})}\int\Omega^a_k(y,{\mathbf b}_1,{\mathbf b}_2)\Omega^a_i(Y-y,{\mathbf b}_1,{\mathbf b}_2)
 d^2b_1d^2b_2\delta^{(2)}({\mathbf b}_1-{\mathbf b}_2-{\mathbf b})\; .
 \label{eq:Nb}
 \end{equation}
where $\sigma_{ik}({\mathbf b})=2[1-\exp(-F_{ik}({\mathbf b})/2)]$ and $a=P_1,P_2,P_3,R$. Recall that $P_1,~P_2$ and $P_3$ are the large, intermediate and small size components of the Pomeron respectively.
After averaging over the impact parameter ${\mathbf b}$ and the diffractive eigenstates
 $i,k$ of the incoming protons, we obtain
\begin{equation}
 N^a_{\rm tube}~=~\frac
 1{\sigma_{\rm tot}\beta_0^2}\sum_{i,k}|a_i|^2|a_k|^2\int
 \Omega^a_k(y,{\mathbf b}_1,{\mathbf b}_2)\Omega^a_i(Y-y,{\mathbf b}_1,{\mathbf b}_2) d^2b_1d^2b_2\; .
  \label{eq:Nc}
 \end{equation}
 This quantity may be considered as the mean number of colour tubes
 of type $a$ produced in the proton-proton interaction. Note that
 the value of $N^a_{\rm tube}$ does not depend\footnote{This
 was checked by straightforward computation.} on the rapidity $y$.

\begin{figure} [t] 
\begin{center}
\includegraphics[height=18cm]{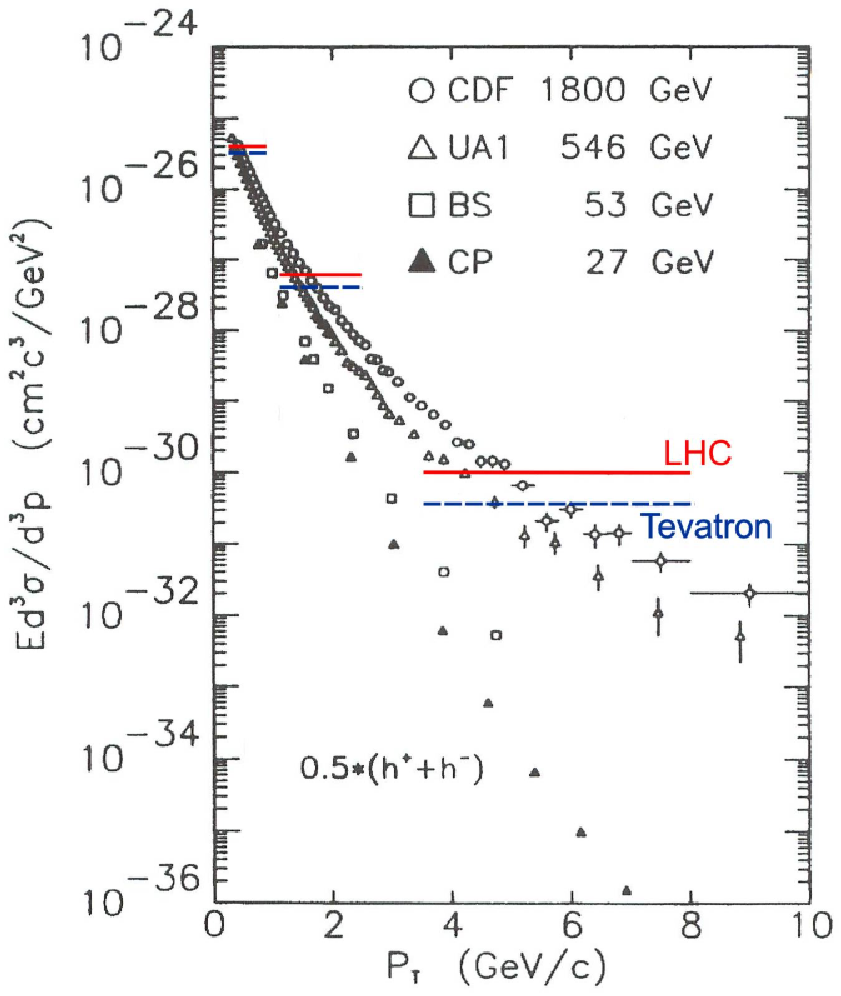}
%\epsf{figure=sumlog.eps,width=14.0cm,height=14.0cm}
%\centerline{\epsfxsize=2.9in\epsfbox{sumlog.eps}}
\caption[*]{The plot is from Ref.~\cite{CDF61}. The horizontal lines, which are superimposed, are our model predictions at the Tevatron and LHC energies; the three $p_t$ ranges correspond to the large-, intermediate- and small-size components of the Pomeron.} 
\label{fig:inclmult}
\end{center}
\end{figure}

 To obtain the number of partons created by the ladder `$a$' at 
 rapidity $y$, we have to include  
 the parton density $\rho^a(y)$ of (\ref{eq:rho}) in the numerator of (\ref{eq:Nc}). That is
\begin{equation}
 N_{\rm parton}^a=\frac
 1{\sigma_{\rm tot}\beta_0^2}\sum_{i,k}|a_i|^2|a_k|^2\int
 \Omega^a_k(y,{\mathbf b}_1,{\mathbf b}_2)\rho^a_{ik}\Omega^a_i(Y-y,{\mathbf b}_1,{\mathbf b}_2) d^2b_1d^2b_2\;
  .  \label{eq:N}
  \end{equation}
The results are shown in Fig.~\ref{fig:sum4}(d).  The main growth in multiplicity, as we go from Tevatron to LHC energies, is due to the small size (`QCD') Pomeron component, which produces particles with typically $p_t \sim 5$ GeV. There is essentially no growth in multiplicity at small $p_t$. This simply confirms the trend that has been observed through the CERN-ISR to Tevatron energy range, see the data points in Fig.~\ref{fig:inclmult}.

In other words, starting with the same intercepts ($\Delta=0.3$) the large size
component contribution after the absorptive correction becomes practically
flat, while  the small size contribution, which is much less affected by
the absorption, continues to grow with energy. As mentioned above, such a behaviour is
consistent with the  experiment (see Fig.~\ref{fig:inclmult}) where the density of a low $k_t$
secondaries is practically saturated while probability to produce a hadron
with a large (say, more then 5 GeV) transverse momentum grows with the
initial energy.

Unfortunately we cannot identify each `parton' with a pion.
At a large $k_t$ it is probable that, after the hadronisation, a parton forms a gluon jet of pions. However, at low $k_t$ this is not evident;
it is hard to say what a gluon `jet' becomes at a low $k_t$.
Nevertheless, in order to compare the model with the data we assume, that 
after hadronisation, each parton from Pomeron components $P_1,\, P_2$ and $P_3$ gives three charged pions
with $p_{ti}\sim 0.5,\, 1.5$ and $5$ GeV respectively. In this way we may estimate the inclusive cross section, at the three values of $p_t$, using
\begin{equation}
\frac{Ed\sigma^{ch}}{d^3p}\ =\ \frac{3\sigma_{\rm tot} N^{P_i}_{\rm parton}}{\pi p_{ti}^2}\ ,
\end{equation}
where $\sigma_{\rm tot}$ allows for normalisation and $1/p_t^2$ accounts for the size of the phase space occupied by the particles from component $i$.
 
To obtain a qualitative feel for the expected behaviour, we show our predictions at the Tevatron and LHC energies in Fig.\ref{fig:inclmult}, where the horizontal lines indicate the typical $p_t$ interval associated with each Pomeron component. Recall, that in our `soft' model, we never use the {\it value} of the Pomeron $k_t$ explicitly. The characteristic parameters actually used in the computations are the {\it ratios} $k^2_{t~i}/k^2_{t~i+1}$. The horizontal lines reflect the $p_t$ intervals covered by the various components arising from the scale choices.  Some features of Fig.\ref{fig:inclmult} are clear. First, although there is some freedom in assigning the overall scale, nevertheless, it appears that the scale choice made in the figure agrees satisfactorily with the Tevatron data \cite{CDF61}. Second, as compared to the Tevatron, the LHC distribution is more enhanced at large $p_t$. The enhancement is a factor of 2.6 for the `QCD' small-size component of the Pomeron, whereas it is only 1.25 for the `soft' Pomeron component.

\section{Summary}

New triple-Regge analyses \cite{LKMR,Pog}, which include absorptive effects, found that the triple-Pomeron coupling is rather large
($g_{3P}= \lambda g_N$ with $\lambda \gapproxeq 0.2$). Thus, in order to obtain reliable predictions for diffractive processes at the LHC, it is necessary to have a model of `soft' high-energy processes which includes multi-Pomeron interactions.

Here we have presented such a model, tuned to the existing `soft' data, which, in principle, is capable of predicting the basic features of high-energy soft $pp$ interactions. 
The absorption of intermediate partons is
described by conventional $\exp(-\lambda\Omega)$-type factors. This corresponds to a coupling
  $g^n_m=nm\lambda^{n+m-2}g_N/2$ of the $n \to m$ Pomeron vertices.

Briefly, the model has multi-components in both the $s$- and $t$-channels. The former are based on a three-channel eikonal approach, together with the inclusion of multi-Pomeron diagrams, so that both low- and high-mass diffractive dissociation are well described. Predictions for the LHC are given. A novel feature of the model is the inclusion of different $t$-channel exchanges, which allows for small-, intermediate- and large-size components of the exchanged Pomeron, each with a {\it bare} intercept $\Delta\equiv\alpha_P(0)-1=0.3$. For the large-size component, the slope of the trajectory is 
$\alpha'_P=0.05$ GeV$^{-2}$. 
The large-size Pomeron component is heavily screened by the effect of `enhanced' multi-Pomeron diagrams, associated with high-mass dissociation. This leads, among other things, to the effective ``saturation'' of the low $p_t$ particle density, and to a slow growth of the total cross section. Indeed, the model predicts a  relatively low total cross section at the LHC -- $\,\sigma_{\rm tot}({\rm LHC})\simeq 90$ mb. On the other hand, the small-size component of the Pomeron is weakly screened, leading to an anticipated growth of the particle multiplicity at large $p_t~~(\sim 5$ GeV) at the LHC. Thus the model has the possibility to embody a smooth matching of the perturbative QCD Pomeron to the `soft' Pomeron.

We emphasized that a reliable model of soft interactions is essential in order to predict the rates of diffractive processes at the LHC. In particular, we used the model to calculate the rapidity gap survival factors, including the effects of {\it both} eikonal and enhanced rescattering.  This is the subject of the following paper \cite{RMK2}.

\section*{Acknowledgements}

We thank Aliosha Kaidalov and Risto Orava for useful discussions.
MGR thanks the IPPP at the University of Durham for hospitality.
The work was supported by  grant RFBR
07-02-00023, by the Russian State grant RSGSS-3628.2008.2.


\begin{thebibliography}{0}

\bibitem{reviews} See, for example, \\
P.D.B. Collins, Regge theory and high energy physics (Cambridge Univ. Press, 1977) \\
  M.M.~Block,
  %``Hadronic forward scattering: Predictions for the Large Hadron Collider and
  %cosmic rays,''
  Phys.\ Rept.\  {\bf 436}, 71 (2006);\\
 %%CITATION = PRPLC,436,71;%
R.~Fiore {\it et al.}, 
%L.~Jenkovszky, R.~Orava, E.~Predazzi, A.~Prokudin and O.~Selyugin,
  %``Forward Physics at the LHC: Elastic Scattering,''
  arXiv:0810.2902 [hep-ph].
  %%CITATION = ARXIV:0810.2902;%%
\bibitem{RFT}V.N.~Gribov, Sov. Phys. JETP {\bf 26}, 414 (1968).
\bibitem{GM1} V.N.~Gribov and A.A.~Migdal, Sov. J. Nucl. Phys. {\bf 8}, 583
(1969).
\bibitem{GM2} V.N.~Gribov and A.A.~Migdal, Sov. Phys. JETP {\bf 28}, 784
(1969).
%\bibitem{RFT}V.N.~Gribov, Sov. Phys. JETP {\bf 26} (1968) 414.
 
\bibitem{MGR}M.~G.~Ryskin, A.~D.~Martin and V.~A.~Khoze,
  %``Diffractive processes at the LHC,''
arXiv:hep-ph/0506272.
%%CITATION = HEP-PH/0506272;%%
 
\bibitem{KMR-08}V.~A.~Khoze, A.~D.~Martin and M.~G.~Ryskin,
  %``Soft Diffraction at the LHC,''
  arXiv:0810.3324 [hep-ph].
  %%CITATION = ARXIV:0810.3324;%%

\bibitem{LKMR} E.G.S.~Luna, V.A.~Khoze, A.D.~Martin and M.G.~Ryskin, Eur. Phys. J. {\bf C59}, 1 (2009).

\bibitem{early} V.A. Khoze, A.D. Martin and M.G. Ryskin, Eur. Phys. J. {\bf C55}, 363 (2008).  

 \bibitem{KMRpr} V.A.~Khoze, A.D.~Martin and M.G.~Ryskin,  Eur. Phys. J.
{\bf C23}, 311 (2002).

\bibitem{KMR} V.A.~Khoze, A.D.~Martin and M.G.~Ryskin, 
Eur. Phys. J. {\bf C14}, 525 (2000). 

\bibitem{DKMOR}A.~De Roeck, V.A.~Khoze, A.D.~Martin, R.~Orava and M.G.~Ryskin,
  %``Ways to detect a light Higgs boson at the LHC,''
  Eur.\ Phys.\ J.\  C {\bf 25}, 391 (2002).
% [arXiv:hep-ph/0207042].
  %%CITATION = EPHJA,C25,391;%%
\bibitem{FP420} M.~Albrow and A.~Rostovtsev,
             arXiv:hep-ph/0009336; \\
  M.~G.~Albrow {\it et al.} [FP420 R and D Collaboration],
  %``The FP420 R&D Project: Higgs and New Physics with forward protons at the
  %LHC,''
  arXiv:0806.0302 [hep-ex].
  %%CITATION = ARXIV:0806.0302;%%
\bibitem{Jz} V.A.~Khoze, A.D.~Martin and M.G.~Ryskin,  Eur. Phys. J. {\bf
C19}, 477 (2001), Erratum {\bf C20}, 599 (2001);
arXiv:hep-ph/0006005.

\bibitem{oldsoft} V.A. Khoze, A.D. Martin and M.G. Ryskin, Eur. Phys. J. {\bf C18}, 167 (2000).

\bibitem{RMK2} M.G. Ryskin, A.D. Martin and V.A. Khoze, arXiv:0812.2413 [hep-ph].

\bibitem{GW} M.L.~Good and W.D.~Walker, Phys. Rev. {\bf 120}, 1857 (1960);\\
E.L. Feinberg and I.Ya. Pomeranchuk, Doklady Akad. Nauk SSSR 
 {\bf 93}, 439 (1953); Suppl. Nuovo Cimento v.~{\bf III}, serie~X, 652 (1956). 

\bibitem{AGK}  V.A.~Abramovsky, V.N.~Gribov and O.V.~Kancheli, Sov. J.
Nucl. Phys. {\bf 18}, 308 (1973).
\bibitem{FF}  R.D.~Field and G.C.~Fox, Nucl. Phys. {\bf B80}, 367 (1974);\\
A.B.~Kaidalov, V.A. Khoze, Yu.F.~Pirogov and N.L.~Ter-Isaakyan, Phys. Lett.
{\bf B45} 471 (1974);\\
for a review see A.B.~Kaidalov, Phys. rep. {\bf 50}, 157 (1979).
\bibitem{capella}A.~Capella, J.~Kaplan and J.~Tran Thanh Van,
  %``Absorptive Corrections To The Inclusive Spectrum And The Bare Triple
  %Pomeron Coupling,''
  Nucl.\ Phys.\  B {\bf 105}, 333 (1976).
  %%CITATION = NUPHA,B105,333;%%
%\mpar{find ref.}
\bibitem{BRV}  J.~Bartels, M.G.~Ryskin and G.P.~Vacca, Eur. Phys. J. {\bf
C27}, 101 (2003).
\bibitem{Jpsi}  ZEUS collaboration: Abstract 549, Int. Europhysics Conf.
on HEP, Aachen, July 2003.
\bibitem{KMRs1} M.G.~Ryskin, A.D.~Martin and V.A.~Khoze,  Eur. Phys. J.
{\bf C54}, 199 (2008).
\bibitem{AFS} D. Amati, A. Stanghellini and S. Fubini, Nouvo Cim. {\bf 26}, 896 (1962).
\bibitem{CDR} E.L. Feinberg and D.S. Chernavski, Usp. Fiz. Nauk {\bf 82}, 41 (1964);\\
         V.N. Gribov, Sov. J. Nucl. Phys. {\bf 9}, 369 (1969);\\
 V.N. Gribov, in {\it Gauge Theories and Quark Confinement}, PHASIS, Moscow, 2002, p.3.
\bibitem{Lip} L.N.~Lipatov, Sov. Phys. JETP {\bf 63}, 904 (1986).

\bibitem{bfkl}   V.S.~Fadin, E.A.~Kuraev, and L.N.~Lipatov,
Phys. Lett. B {\bf60}, 50  (1975); \\
E.A.~Kuraev, L.N.~Lipatov, and V.S.~Fadin, Zh. Eksp. Teor. Fiz.
{\bf 71}, 840 (1976) [Sov. Phys. JETP {\bf 44}, 443 (1976)]; {\it
ibid.} {\bf 72}, 377 (1977) [{\bf 45}, 199 (1977)];\\
I.I.~Balitsky and L.N.~Lipatov, Yad. Fiz. {\bf28}, 1597 (1978)
[Sov. J. Nucl. Phys. {\bf28}, 822 (1978)].

\bibitem{CERN-ISR} L.~Baksay {\it et al.}, Phys.\ Lett.\ {\bf B53} 484 (1975); \\
R.~Webb {\it et al.}, Phys.\ Lett.\ {\bf B55} 331 (1975); \\
L.~Baksay {\it et al.}, Phys.\ Lett.\ {\bf B61} 405 (1976); \\
H.~de Kerret {\it et al.}, Phys.\ Lett.\ {\bf B63} 477 (1976); \\
G.C.~Mantovani {\it et al.}, Phys.\ Lett.\ {\bf B64} 471 (1976).

\bibitem{FSC} M.~Albrow {\it et al.},
  %``Forward Physics with Rapidity Gaps at the LHC,''
arXiv:0811.0120 [hep-ex].
  %%CITATION = ARXIV:0811.0120;%% 
  \bibitem{BFKLnnl}
V.S.~Fadin and L.N.~Lipatov, Phys. Lett. {\bf B429}, 127 (1998);\\
G.~Camici and M.~Ciafaloni, Phys. Lett. {\bf B430}, 349 (1998);\\
G.P.~Salam, JHEP {\bf 9807}, 019 (1998), Act. Phys. Pol. {\bf B30}, 3679
(1999);\\
M.~Ciafaloni, D.~Colferai and G.P.~Salam, Phys. Lett. {\bf B452}, 372
(1999), Phys. Rev. {\bf D60}, 114036 (1999). 

\bibitem{KGB} S. Sapeta and K.J. Golec-Biernat, Phys. Lett. {\bf B613}, 154 (2005). 

\bibitem{GLMM} E.~Gotsman, E.~Levin, U.~Maor and J.S.~Miller,
arXiv:0805.2799.

\bibitem{CDFhm} F. Abe {\it et al.}, [CDF collaboration]  Phys. Rev. {\bf D50} 5535 (1994).

\bibitem{GoulMon} K. Goulianos and J. Montanha, Phys. Rev. {\bf D59} 114017 (1999).

\bibitem{DHS} R. Dolen, D. Horn and C. Schmid, Phys. Rev. {\bf 166}, 1768 (1968). 
%\bibitem{oldsoft} V.A. Khoze, A.D. Martin and M.G. Ryskin, Eur. Phys. J. {\bf C18} (2000) 167.

\bibitem{CDF61}CDF Collaboration, F. Abe et al., Phys. Rev. Lett. {\bf 61}, 1819 (1988).

\bibitem{Pog} T. Poghosyan and A.B. Kaidalov, private communication.

\end{thebibliography}
\end{document}